\documentclass[12pt]{iopart}

\usepackage{hyperref}
\hypersetup{
	colorlinks=true,
	linkcolor=blue,
	filecolor=blue,      
	urlcolor=blue,
	citecolor=blue,
}
\usepackage{graphicx}  
\usepackage{subfig}
\usepackage[utf8]{inputenc}
\usepackage{cite}
\usepackage{multirow}
\usepackage{makecell}
\usepackage{array}
\usepackage{inputenc}
\DeclareUnicodeCharacter{2212}{\textendash}
\begin{document}

\title[]{Theoretical studies on structural properties and decay modes of $^{284-375}$119 isotopes}
\author{Asloob A. Rather$^{*,1,2}$, M. Ikram$^{3}$, Ishfaq A. Rather$^{4}$, M. Imran$^1$,
	A. A. Usmani$^1$, Bharat Kumar$^5$, K. P. Santhosh$^{6,7}$, S. K. Patra$^8$ }
\address{$^1$Department of Physics, Aligarh Muslim University, Aligarh 202002, India.}
\address{$^2$Department of School Education, Govt of Union Territory of Jammu and Kashmir, India.}
\address{$^3$ Department of Physics, Harsh Vidya Mandir (PG) College Raisi, Haridwar-247671, Uttarakhand, India. } 
\address{$^4$ Centro de Astrof{\'i}sica e Gravita{\c c}{\~a}o-CENTRA, Instituto Superior T{\'e}cnico-IST, Universidade de Lisboa-UL,
	Av.~Rovisco Pais, 1049-001 Lisboa, Portugal.}  
\address{$^5$Department of Physics and Astronomy, NIT Rourkela-- 769008, Odisha, India.}
\address{$^6$School of Pure and Applied Physics, Kannur University, Payyanur Campus, Payyanur-670327, Kerala, India.}
\address{$^7$Department of Physics, University of Calicut, Kerala-673635, India.}
\address{$^8$Institute of Physics, Sachivalaya Marg, Bhubaneswar 751005, India.}
\ead{$^*$asloobamu@gmail.com,$^*$asloobamu@yahoo.com}
\vspace{10pt}

\begin{abstract}
In this manuscript, we analyze the structural properties of $Z=119$ superheavy 
nuclei in the mass range of 284 $\le$ A $\le$ 375 within the framework of 
axially deformed relativistic mean field theory (RMF) 
and calculate the binding energy, radii, quadrupole deformation parameter,
separation energies and density profile.
To investigate the phenomenon of shape coexistence the RMF calculations are 
performed within three possible solutions i.e. prolate, oblate and
spherical configurations. To get a better visualization of nucleon and total
matter distribution, two-dimensional contour representation of density 
distribution for $^{291}$119 and $^{303}$119 has been made.
Further, a competition between possible decay modes such as $\alpha-$decay, 
$\beta-$decay and spontaneous fission (SF) of the isotopic chain of $Z=119$
superheavy nuclei under study is systematically analyzed 
within self-consistent relativistic mean field model.
The $\alpha$-decay half lives are estimated using the semi-empirical
formulae by Viola-Seaborg 
[ J. Inorg. Nucl. Chem. {\bf28}, 741 (1966).], B. A. Brown 
[ Phys. Rev. C {\bf46}, 811 (1992).], G. Royer 
[ J. Phys. G, Nucl. Part. Phys. {\bf26}, 1149 (2000).], 
N. Dasgupta-Schubert and M.A. Reyes 
[ At. Data Nucl. Data Tables {\bf93}, 90 (2007).]  
D. D. Ni et al.,[ Phys. Rev. C, {\bf78}, 044310 (2008).] and 
a close agreement is noticed amongst these and also with the estimations
made by Finite Range Droplet Model (FRDM) wherever available.
Moreover, our analysis confirmed that $\alpha-$decay is restricted
within the mass range 284 $\leq$ A $\leq$ 296 and thus being the dominant
decay channel in this mass range. 
There is no possibility of $\beta-$decay for the considered isotopic chain. 
In addition, we forecasted the $\alpha-$decay chain of fission survival nuclides 
i.e. $^{284-296}$119 and found as one $\alpha$ chain from $^{284}$119 and $^{296}$119,
two consistent $\alpha$ chains from $^{285}$119 and $^{295}$119,
three consistent $\alpha$ chains from $^{286}$119 and $^{294}$119,
four consistent alpha chains from $^{287}$119, six consistent alpha
chains from $^{288-293}$119. 
Further, SF half lives and alpha half-lives are computed using the formula of Santhosh et al and Coulomb Proximity Potential model (CPPM) for the $^{284-296}$119 isotopic chain. It has been observed that the isotopes $^{288-293}$119 exhibit 6$\alpha$ chains followed by SF, isotopes $^{295,296}$119 exhibit 4$\alpha$ chains followed by SF and the rest of the nuclei show continuos $\alpha$ chains. This study has also established that the alpha half-life values computed using Q$_\alpha$ (RMF) agree with half-life values computed using experimental Q$_\alpha$ values within 1 order difference. 
Thus, such studies can be of great significance to
the experimentalists in very near future for synthesizing $Z=119$
superheavy nuclei.	
\end{abstract}

%

\section{Introduction}
\label{intro}
Theoretical and experimental studies of the nuclei 
with large number of neutrons and protons has
witnessed an upsurge and has become the 
subject of intense debate among nuclear physics community
from past several decades. 
Thus, exploring the existence limit of very heavy nuclei, i.e., nuclei 
with Z~$\geq$~104 and island of stability in superheavy nuclei (SHN) 
has been a challenging issue in nuclear physics 
from a fairly long period of time. 
The discovery of new superheavy elements (SHEs) has lead to the simultaneous 
expansion of periodic table and Segre chart of nuclei. 
Hence, the studies based on the identification of new SHN would 
extend our knowledge about the nuclear potentials and resulting nuclear structure. 
The hunt for SHN started in the late 1960s with the island 
of stability around $Z=114$ and $N=184$~\cite{CDHMN96}. 
The existence of superheavy nuclei is the result of the interplay 
between large disruptive coulomb force and the attractive nuclear potential. 
Owing to the large number of protons in SHN the Coulomb disruption dominates
the attract nuclear force thus making the SHN unstable
and therefore highly susceptible to spontaneous fission. The
question that arise then is what makes these SHN stable.
The answer to this question came however
by the end of 1960s, when it was firmly established that the existence
of heavier nuclei with Z~$\geq$~104 was primarily determined by the 
quantum mechanical shell effects i.e. single-particle motion of 
neutrons and protons in quantum orbits~\cite{MS66,SGK66,MG69,CPDN83,PS91}.
The next fundamental question that nuclear physics community 
try to find out is the maximum possible combination of neutrons and proton 
that can found or synthesized in the laboratory.
With the huge progress in theory, experiments and accelerator technologies 
and the advent of  state-of-art radioactive ion beam facilities, it has 
become possible to synthesize the superheavy nuclei and 
reach to the island of stability in superheavy nuclei. 
The process of synthesizing SHN is done via fusion evaporation reactions, i.e. 
cold fusion reaction~\cite{HM0} and hot fusion reaction~\cite{O07}. 
The cold fusion technique which involves doubly magic spherical 
target and deformed projectile has been successful in synthesizing 
$Z=107-113$~\cite{HACK04,A07,M15} at GSI, Darmstadt and RIKEN Japan. 
On the other hand hot fusion reaction using neutron-rich $^{48}$Ca beams on 
actinide targets, the synthesis of $Z=107-118$ have been done at 
JINR-FLNR, Dubna~\cite{OU15}. 
Recently in 2009, an attempt to synthesis $Z=120$ by using hot 
fusion reaction was made by Oganessian et al.~\cite{OUR09}. 
However, due to the low cross-section values of the order of picobarn
and sub-picobarn levels obtained in the experiments for
synthesizing SHN makes the experiment to last for several months
and henceforth results in the identification of few events(nuclei).
Analysis of low-statistics data and investigation of new isotopes 
becomes of crucial importance. 
Thus, running of experiments for long periods results in optimization 
of production methods through the determination of excitation functions 
as demonstrated in  recent studies of the $^{232}$Am+ $^{48}$Ca 
reaction~\cite{YTO12,YTOE7}.

The elusive superheavy mass region provide an opportunity
to nuclear physicists to explore the concepts like magic numbers and
island of stability, which help us to understand why certain nuclei are
more stable than others. 
Various theoretical investigations have been carried using microscopic-macroscopic
approaches and the self-consistent mean-field in both the relativistic and 
non-relativistic domains~\cite{PXLZ07,SFM05} and the primary goal of these 
studies is to find out the combination of neutrons and protons where 
spherical shell closure may occur. However, there is no general consensus
among relativistic and non-relativistic theoretical models in predicting 
spherical shell closures. For instance, the nuclear shell model predicts
the next magic number beyond $Z=82$ at $Z=114$. However, the microscopic-macroscopic
model predicts it to be at $Z=114$ and $N=184$~\cite{MS67,M67,NTSSWGLMN69} which is 
considered to be the island of superheavy mass region and confirmation 
of it has become a much debated issue nowadays. 
There is no confirmation till date regarding the center of island of stability in SHN. 
Analogues to mic-mac predictions, the microscopic models predicts closed 
spherical shell closures at $N=184$ but for nuclei with higher number 
of protons, i.e., $Z=120,122,124$ or 126~\cite{CDHMN96,BRRMG98,RBBSRMG97,tsil2004}. 
However, it is to be noted that most of the theoretical investigations
predict $N=184$ as the neutron magic number. The fragility/uncertainty 
in predicting the correct proton magic number is attributed to the ambiguous
strength of spin orbit coupling which posses a great difficulty in 
localization of single-particle energy levels  between $Z=114$ and 126.

The theoretical investigations carried out specifically on $\alpha-$decay
properties in superheavy mass have close connection to the nuclear model
predictions~\cite{WNH99,DZS11,PGG11} like clustering, shell structures, 
deformations and quasi-particle excitations and various theoretical 
approaches have been put forth for computation
of $\alpha-$decay properties in the superheavy mass region.
One of the effective ways possibly to study SHNs is via the characterizations 
of their decay properties and in particular $\alpha-$decay is considered to be
an inevitable tool to identify and study SHN as it provides world of
information regarding the nuclear structure. The prominent mode of decay
in superheavy mass region is alpha decay followed by spontaneous fission.
The proper measurement of alpha decay properties provide useful inputs
on structure of superheavy nuclei, for instance, shell effects and stability,
nuclear spins and parities, deformation, rotational properties, fission barrier, etc.
The credit to the discovery of $\alpha-$decay goes to Rutherford~\cite{RG08,RR09}
in 1899 and Gamow~\cite{G28} was first to describe it in 1928 using
the concept of quantum tunnelling through potential barrier.
Currently various theoretical investigations  which  belong
to macro-micro methods like the cluster model~\cite{BMP92}, 
fission model~\cite{PISG85}, the density dependent M3Y (DDM3Y)
effective model~\cite{B03}, the generalized liquid drop model (GLDM)
~\cite{ZG07}, etc and the self-consistent models like
relativistic mean-field theory~\cite{SFM05}, Skyrme-Hartree-Fock mean
field model~\cite{WWHG15} are being employed to explain the $\alpha-$decay 
from heavy and superheavy nuclei. Recently, working within the ambit of axially
deformed relativistic mean field model by employing NL3$^{*}$ parameterization, 
a systematic study of alpha decay half lives of predicted
magic nuclei $Z=132$, 138~\cite{Asloob2016} in the mass range 
312 $\leq$ A $\leq$ 392 has been made and computation of alpha 
decay half lives was performed by using the semi-empirical formulae
VSS~\cite{VJS66}, Brown~\cite{B92}, Royer~\cite{R0}, GLDM~\cite{DR07} 
and Ni et al.,~\cite{NDK08}. 
By employing 20 mass models and 18 empirical formulae an extensive 
and systematic study was performed by Wang et al.~\cite{WLWM14} on 
alpha decay energies and alpha decay half lives of superheavy 
nuclei with Z~$\geq$~100 respectively and established
that for reproducing the Q$_\alpha$ values of SHN, the WS4 mass model is most
appropriate one. Moreover, the outcome of these studies firmly authorized that out
of 18 empirical formulae SemFIS2~\cite{PGC07} is the most reliable 
one to predict alpha
decay half lives as the parameters involved in the formula are taken from
experimental alpha emitter data of transuranium nuclei including SHN ($Z=92-118$)
and the UNIV2~\cite{PGC07} formulae with fewest parameters
is also effective in superheavy mass region.
Moreover, VSS~\cite{VS66,SPC89}, SP~\cite{SP07,PS05} and NRDX~\cite{NRDX08} employing 
fewer parameters  are also very handy in the prediction of alpha decay half lives.

Although both alpha decay and spontaneous fission are explained by quantum 
mechanical tunnelling, the two widely differ in principle. 
Whereas alpha decay is described as the alpha cluster penetrating the coulomb 
barrier after its formation in the parent nucleus, the process of spontaneous 
fission is much more intricate as it involves large uncertainties such as mass 
and charge numbers of the two fragments, the number of emitted neutrons, and the 
released energy etc. It is to be emphasized that though alpha decay and 
spontaneous fission are the principal modes of decay of superheavy nuclei 
with Z$~\geq$~92, it the spontaneous fission that acts as limiting factor 
for determining the stability of superheavy nuclei. 
In 1939 Bohr and Wheeler~\cite{BW39} described the mechanism of spontaneous 
fission and established a limit $\frac{Z^{2}}{A}\approx 48$
for SF beyond which nuclei are susceptible to spontaneous fission. 
Flerov et al.~\cite{FP40} observed SF from $^{238}$U and this was followed 
by several empirical formulae being proposed for determining
the SF half lives and it was Swatecki~\cite{S55} in 1955 who 
put forward the first semi-empirical formulae for estimation of SF half lives. 
Presently, we come across the globe in different laboratories~\cite{O5,O6,G06,D06,P06,O4} 
SF half lives being measured and extensive theoretical investigations 
carried out by several theoretical groups for identifying the long 
lived superheavy elements. 
Several empirical formulae have been proposed for estimation of SF fission 
half lives by different researchers. 
Xu et al.~\cite{XRG08} put forward a semi-empirical formula for estimating SF half 
life of even-even nuclei using parabolic potential and the agreement between
theoretical and experimental results is quite good. A phenomenological
formula proposed by Ren et al.~\cite{RX05,XR05} in 2005 for calculating
SF half lives of even-even nuclei was generalized to both the case of odd nuclei
and fission isomers. Within the microscopic-macroscopic model approach, 
Smolanczuk et al.~\cite{SSS95} calculated the SF properties for deformed even-even,
odd-A and odd-odd superheavy nuclei with $Z=104-120$. This was followed by
computation of spontaneous fission barriers of $Z=96-120$ by
Muntain et al.~\cite{MPS03} within microscopic-macroscopic model. By employing
Hartree-Fock-Bogoliubov(HFB) approach with finite range
and density dependent Gogny force with the DIS parameter set
Warda et al.~\cite{WE12} estimated the SF half lives of 160
heavy and superheavy nuclei. The study carried out by
Stasczak et al.~\cite{SBN13} by using density functional
theory for estimation of SF half lives and life times of
superheavy elements presented a systematic self-consistent approach to SF in SHN. 
The computation of SF half lives using the semi-empirical formula 
by Ren and Xu for $Z=132$, 138 with mass ranges 312 $\leq$ A $\leq$ 392 and 
318 $\leq$ A $\leq$ 398 has been done recently and reported in Ref.~\cite{Asloob2016}.
Here, in present manuscript we made an attempt to analyze the 
competition among various possible modes of decay of $Z=119$ superheavy 
nuclei such as $\alpha-$decay, $\beta-$decay and SF along with 
the structural studies and predict the principal
mode of decay of considered isotopic chain.
Further, we performed the study about feasibility of observing
the $\alpha-$decay chains for fission survival nuclides i.e. 
$^{284-297}$119 of the considered isotopic chain.
The contents of the manuscript are organized as follows: 
The framework of relativistic mean-field formalism is outlined in section two.
Results and discussion is presented in section three.
Finally, section four contains the main summary and conclusions of this work. 
\section{Theory and Formalism}
\label{sec:headings}
\subsection{Axially Deformed Relativistic Mean-Field}
From last few decades, the relativistic mean field theory has been 
successfully reproduced the ground state energy and other 
physical observables of the nuclei throughout the periodic table 
near as well as far from the stability line including superheavy 
valley~\cite{pannert1987,S92,GRT90,R96,SW86,BB77,crp1991,ren2003,sharma2005}.
The starting point of the RMF theory is the basic Lagrangian density 
containing nucleons interacting via exchange of $\sigma-$, $\omega-$ and $\rho-$mesons. 
The contribution of $\pi-$meson is zero at mean field due to its pseudo scalar nature. 
Thus, $\sigma-$, $\omega-$ and $\rho-$ are only the mesonic field in 
which $\sigma-$, $\omega-$ mesons reproduce the large scalar and vector 
potentials and as a result originate the reasonable nuclear mean potential 
and large spin-orbit potential. The $\rho-$meson takes the care of 
nuclear asymmetry of the systems. Moreover, photon field $A_{\mu}$ is 
included to handle the Coulomb interaction between protons. 
The relativistic mean field Lagrangian density is 
expressed as~\cite{pannert1987,S92,GRT90,R96,SW86,BB77},
\begin{eqnarray}
{\cal L}&=&\bar{\psi_{i}}\{i\gamma^{\mu}
\partial_{\mu}-M\}\psi_{i}
+{\frac12}\partial^{\mu}\sigma\partial_{\mu}\sigma
-{\frac12}m_{\sigma}^{2}\sigma^{2}
-{\frac13}g_{2}\sigma^{3} \nonumber \\
&-&{\frac14}g_{3}\sigma^{4}-g_{\sigma}\bar{\psi_{i}}\psi_{i}\sigma 
-{\frac14}\Omega^{\mu\nu}\Omega_{\mu\nu}+{\frac12}m_{w}^{2}V^{\mu}V_{\mu}\nonumber \\
&-&g_{w}\bar\psi_{i}\gamma^{\mu}\psi_{i}
V_{\mu}-{\frac14}\vec{B}^{\mu\nu}\vec{B}_{\mu\nu} 
+{\frac12}m_{\rho}^{2}{\vec{R}^{\mu}}{\vec{R}_{\mu}}-{\frac14}F^{\mu\nu}F_{\mu\nu} \nonumber\\
&-&g_{\rho}\bar\psi_{i}\gamma^{\mu}\vec{\tau}\psi_{i}\vec{R^{\mu}}-e\bar\psi_{i}
\gamma^{\mu}\frac{\left(1-\tau_{3i}\right)}{2}\psi_{i}A_{\mu} .
\end{eqnarray}
Here M, $m_{\sigma}$, $m_{\omega}$ and $m_{\rho}$ are the masses for nucleon, 
${\sigma}-$, ${\omega}-$ and ${\rho}-$mesons and ${\psi}$ is its Dirac spinor. 
The field for the ${\sigma}-$meson is denoted by ${\sigma}$, ${\omega}-$meson 
by $V_{\mu}$ and ${\rho}-$meson by $R_{\mu}$. 
The quantities $g_\sigma$, $g_{\omega}$, $g_{\rho}$ and $e^2/4{\pi}$=1/137 
are the coupling constants for the ${\sigma}-$, ${\omega}-$, ${\rho}-$mesons and 
photon field respectively. The $g_2$ and $g_3$ are the nonlinear self-interaction 
coupling constants for ${\sigma}-$mesons. By using the classical variational principle,
we obtain the field equations for the nucleons and mesons known by Dirac and 
Klein-Gordon equations.  
The Dirac equation for the nucleons is written by
\begin{equation}
\{-i\alpha\bigtriangledown + V(r_{\perp},z)+\beta M^\dagger\}\psi_i=\epsilon_i\psi_i.
\end{equation}
The effective mass of the nucleon is
\begin{equation}
M^\dagger=M+S(r_{\perp},z)=M+g_\sigma\sigma(r_{\perp},z),
\end{equation}
and the vector potential is
\begin{equation}
V(r_{\perp},z)=g_{\omega}V^{0}(r_{\perp},z)+g_{\rho}\tau_{3}R^{0}(r_{\perp},z)+
e\frac{(1-\tau_3)}{2}A^0(r_{\perp},z). 
\end{equation}
Further, the Klein-Gordon equations are written like as 
\begin{eqnarray}
\{-\bigtriangleup+m^2_\sigma\}\sigma^0(r_{\perp},z)&=&-g_\sigma\rho_s(r_{\perp},z)-g_2\sigma^2(r_{\perp},z)\nonumber\\
&&- g_3\sigma^3(r_{\perp},z) ,\\
\{-\bigtriangleup+m^2_\omega\}V^0(r_{\perp},z)&=&g_{\omega}\rho_v(r_{\perp},z) ,\\
\{-\bigtriangleup+m^2_\rho\}R^0(r_{\perp},z)&=&g_{\rho}\rho_3(r_{\perp},z) , \\
-\bigtriangleup A^0(r_{\perp},z)&=&e\rho_c(r_{\perp},z). 
\end{eqnarray}
Here $\rho_s(r_{\perp},z)$, and $\rho_v(r_{\perp},z)$ are 
the scalar and vector density for $\sigma-$ and $\omega-$fields 
in nuclear system which are expressed as
\begin{eqnarray} 
\rho_s(r_{\perp},z) &=& \sum_ {i=n,p}\bar\psi_i(r)\psi_i(r)\;,                         \nonumber \\
\rho_v(r_{\perp},z) &=& \sum_{i=n,p}\psi^\dag_i(r)\psi_i(r) \;.			      
\end{eqnarray}
The vector density $\rho_3(r_{\perp},z)$ for $\rho-$field and charge density 
$\rho_c(r_{\perp},z)$ are expressed by 
\begin{eqnarray} 
\rho_3(r_{\perp},z) &=& \sum_{i=n,p} \psi^\dag_i(r)\gamma^0\tau_{3i}\psi_i(r)\; ,		      \nonumber \\
\rho_c(r_{\perp},z) &=& \sum_{i=n,p} \psi^\dag_i(r)\gamma^0\frac{(1-\tau_{3i})}{2}\psi_i(r)\;.    
\end{eqnarray}
A static solution is obtained from the equations of motion to describe 
the ground state properties of nuclei.
The set of nonlinear coupled equations are solved self-consistently
in an axially deformed harmonic oscillator basis $N_F=N_B=20$ and 
we obtain all the physical observables. 
The quadrupole deformaton parameter $\beta_2$ is extracted 
from the calculated quadrupole moments of neutrons and protons through 
\begin{equation}
Q=Q_n+Q_p=\sqrt{\frac{16\pi}{5}}(\frac{3}{4\pi}AR^2\beta_2),
\end{equation}
where $R=1.2A^{1/3}$.\\
The various rms radii are defined as
\begin{eqnarray}
\langle r_p^2\rangle &=& \frac{1}{Z}\int r_p^{2}d^{3}r\rho_p(r_{\perp},z)\;,        \nonumber \\
\langle r_n^2\rangle &=& \frac{1}{N}\int r_n^{2}d^{3}r\rho_n(r_{\perp},z)\;,        \nonumber \\
\langle r_m^2\rangle &=& \frac{1}{A}\int r_m^{2}d^{3}r\rho(r_{\perp},z)\;,          
\end{eqnarray}
for proton, neutron and matter rms radii, respectively.
The quantities $\rho_p(r_{\perp},z)$, $\rho_n(r_{\perp},z)$ and $\rho(r_{\perp},z)$ are their 
corresponding densities. 
The charge rms radius can be found from the proton rms radius 
using the relation $r_{c} = \sqrt{r_p^2+0.64}$ by taking finite size of proton into 
consideration.
The total energy of the system is given by 
\begin{equation}
E_{total} = E_{part}+E_{\sigma}+E_{\omega}+E_{\rho}+E_{c}+E_{pair}+E_{c.m.},
\end{equation}
where $E_{part}$ is the sum of the single particle energies of the nucleons and 
$E_{\sigma}$, $E_{\omega}$, $E_{\rho}$, $E_{c}$, $E_{pair}$, $E_{cm}$ are 
the contributions of the meson fields, the Coulomb field, pairing energy  
and the center-of-mass energy, respectively. In present calculations, we use the 
constant gap BCS approximation to take care of pairing interaction~\cite{madland}.
The non-linear NL3* parameter set~\cite{LKFAAR09} is used throughout the calculations.
\subsection{Coulomb-Proximity Potential Model(CPPM)}
The Coulomb-Proximity Potential Model (CPPM)~\cite{SJ00,SBSJ08} was introduced to study alpha and cluster radioactivity and proved as an effective tool in explaining alpha decay of nuclei in superheavy region ~\cite{SBS09,SSB09}. The total potential in CPPM is taken as a sum of Coulomb interaction between daughter nuclei with charge number Z$_1$ and alpha particle with charge number Z$_2$; nuclear proximity potential, VP(z); and centrifugal potential given as: 
\begin{equation}
V = \frac{Z_1Z_2e^2} {r} + V_p(z)+ \frac{ \hbar^2 l (l+1)}{ 2 \mu r^2}
\end{equation}
 Here ‘r’ is the distance between the fragment centers, ‘z’ is the distance between the near surfaces of the fragments, $‘l’$ is the angular momentum and ‘$\mu$’ is the reduced mass.
 \begin{equation}
 V_P(z)= 4 \pi \gamma b  \left[ \frac{C_1 C_2}{C_1 + C_2}\right]  \Phi \left( \frac{z}{b}\right) 
 \end{equation}
 is the proximity potential~\cite{BRST77} with $\gamma$  the nuclear surface tension coefficient and $\Phi$ the universal proximity potential~\cite{BS81}. Here $b \approx 1$ fermi is the diffuseness of the nuclear surface, and  C$_i$  (i=1,2) are the Süsmann central radii of the fragments.
 The barrier penetrability P is given as, 
 \begin{equation}
 P = exp\left\lbrace - \frac{2}{\hbar} {\int_{a}^{b} \sqrt{2 \mu \left( V-Q\right) } } dz
  \right\rbrace 
 \end{equation}
Using the condition, V (a) = V (b) = Q, where Q is the energy released, the turning points a and b can be determined. The decay half-life, $T_ {1/2} = \frac{ln2}{\lambda} = \frac{ln2}{\nu P}$  where $\lambda$  is decay constant and the assault frequency, $\nu = \frac{\omega}{2\pi} = \frac{2E_\nu}{h}$. The empirical vibration energy E$_\nu$, is given as~\cite{PISG85},
\begin{equation}
E_\nu = Q \left\lbrace  0.056 + 0.039 exp \left[ \frac{\left(4-A_2\right) }{2.5}  \right]   \right\rbrace for A_2 \geq 4
\end{equation} 
\section{Results and discussions}
It is worth mentioning that till now the superheavy nuclei up 
to $Z=118$~\cite{ogi2006,ogi2007} have been synthesized in the 
laboratory and experiments have also been attempted 
for the production of $Z=120$~\cite{OUR09}, however its production 
cross-section is very small. Thus, it desires to choose a proper combination of 
projectile and target in hot fusion reaction to improve the production 
cross-section for magic proton shell nuclei (i.e. $Z=120$). On theoretical 
estimation of evaporation residue cross-section, many of the possibilities of 
hot fusion reactions are suggested regarding the synthesizaton of nuclei with 
$Z=120$~\cite{liu2009,wang2012,zhu2014,ansari2016}.
Not only this, evaporation residue cross-section for superheavy nuclei with 
$Z=119$ has also been predicted by number of 
people~\cite{wang2012,zhu2014,ansari2016,adamian2018} and 
found that this nucleus might be produced easier than the magic proton 
shell nuclei~\cite{guo2011}.
Therefore, experiment to produce isotopes of $Z=119$ using hot 
fusion reactions is of great interest and it would bridge the gap between 
experimentally known $Z=118$ and magic proton shell nuclei. 
Regarding the observation of SHN, it is noticed that the superheavy 
nuclei are identified by $\alpha-$decay in the
laboratory followed by spontaneous fission.
In this view, it make sense to have some theoretical predictions on decay 
channels of $Z=119$ superheavy nuclei for guiding the experiment.
Concerning to this, we make mean field calculations to analyze the 
competition among $\alpha-$decay, $\beta-$decay and spontaneous 
fission for predicting the possible mode of decay of isotopic chain 
under study and this is considered to be central theme of the paper.   
In addition to gain some structural information, we calculate the 
total binding energy (BE), radii, quadrupole deformation parameter 
($\beta_{2}$) and density profile for three possible
shape configurations in the mass range of
284 $\le$ A $\le$ 375 which covers many of the neutron magic numbers.
The results concerning to structure and decay of $Z=119$ isotopic chain 
are fully explained in subsections 3.1 to 3.5.
\subsection{Binding energy, radii and quadrupole deformation parameter}
The calculated binding energy, radii and quadrupole deformation parameter 
for the isotopic chain $^{284-375}119$ are given in Tables 1, 2 and 
plotted in Figures 1, 2. 
To identify the possible ground state configuration of the nuclei, 
the field equations are solved with an initial spherical, prolate and oblate quadrupole 
deformation parameter $\beta_{0}$ in relativistic mean field formalism.
Nucleus, a quantum many body system,  acquires different binding energy by 
their possible shape configurations leading to the ground as well as 
intrinsic excited states. It is worthy to mention that maximum binding 
energy of a quantum system corresponds to the ground state energy of the 
system and all other solutions may correspond to the intrinsic excited states. 
Concerning these facts into consideration, we found prolate as a ground state 
for most of the nuclides of $Z=119$. Thus, structural properties 
and decay energies are plotted and estimated for prolate shaped throughout 
the chain. Moreover, some nuclides do not have all three well defined 
shape and we obtain only two solutions of the field equations. 
As the experimental informations of these isotopes are not available, so
in order to provide some validity to the predictive power of our model and their results
a comparison of binding energies of our calculations with those obtained
from finite range droplet model (FRDM)~\cite{moller} is made 
wherever available and some how close agreement is found among them. 
From Table 1, we can see that the binding energies difference between RMF 
and FRDM is very small. The maximum difference between RMF and FRDM values is about 7 MeV, 
namely, the relative differences are less than 0.35\%.
Our calculated one- and two-neutron separation energy is also matches well 
with FRDM estimations.
However, there is no agreement in quadrupole deformation parameter 
within RMF and the values obtained from FRDM data~\cite{moller}. 
Some of the nuclides of considered isotopic series, for example $^{298-318}$119 
having large prolate quadrupole deformation parameter and therefore supposed to 
be superdeformed by their shape. 
Superdeformation is common phenomenon in RMF calculations and it plays a significant 
role for stability of superheavy nuclei.
The radii increases with increasing the mass number and a sudden change 
in radii indicates the change in shape of the nuclides. 
In general, the calculated binding and separation energies from RMF are in good 
agreement with those of the FRDM values wherever available.
\begin{figure}
\centering
	\resizebox{0.70\textwidth}{!}{
		\includegraphics{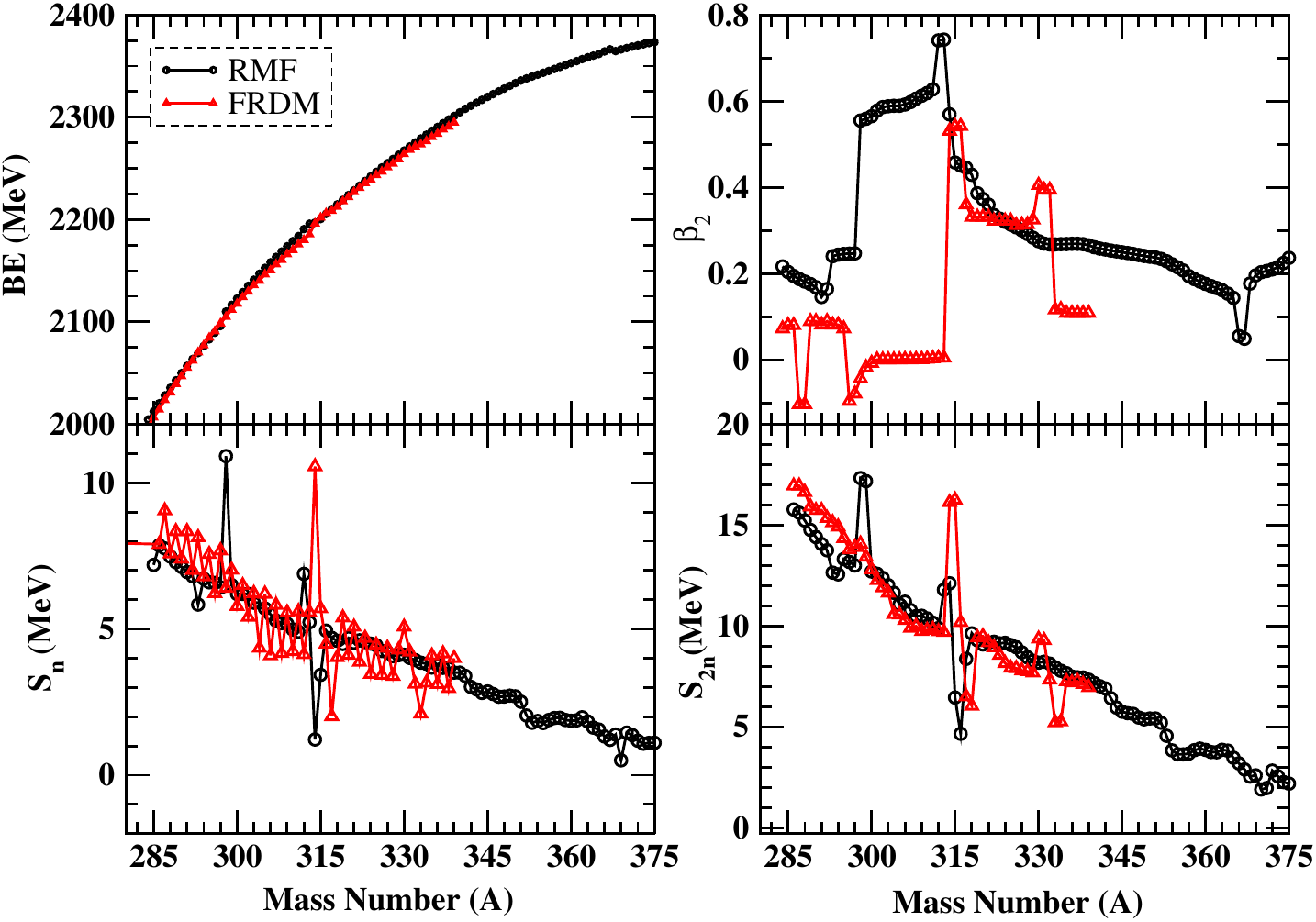}
	}
	\caption{(color online) Binding energy, quadrupole deformation parameter, one 
		and two-neutron separation energy are given as function of mass number.}
	\label{structure}
\end{figure}

\begin{figure}
	\centering
	\resizebox{0.70\textwidth}{!}{%
		\includegraphics{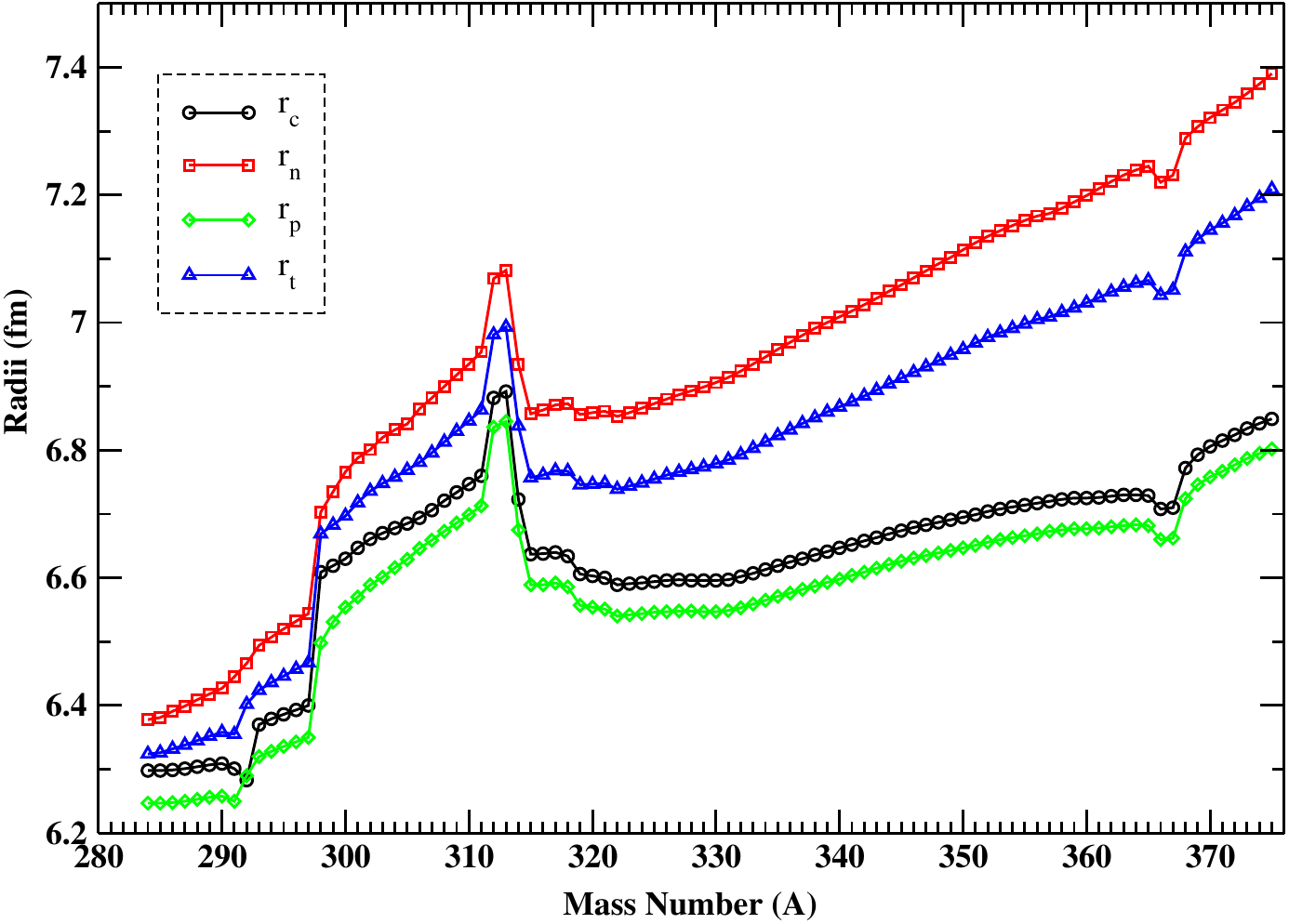}
	}
	\caption{(color online) Radii as a function of mass number.}
	\label{radii}
\end{figure}
\begin{figure}
\centering
	\resizebox{0.70\textwidth}{!}{%
		\includegraphics{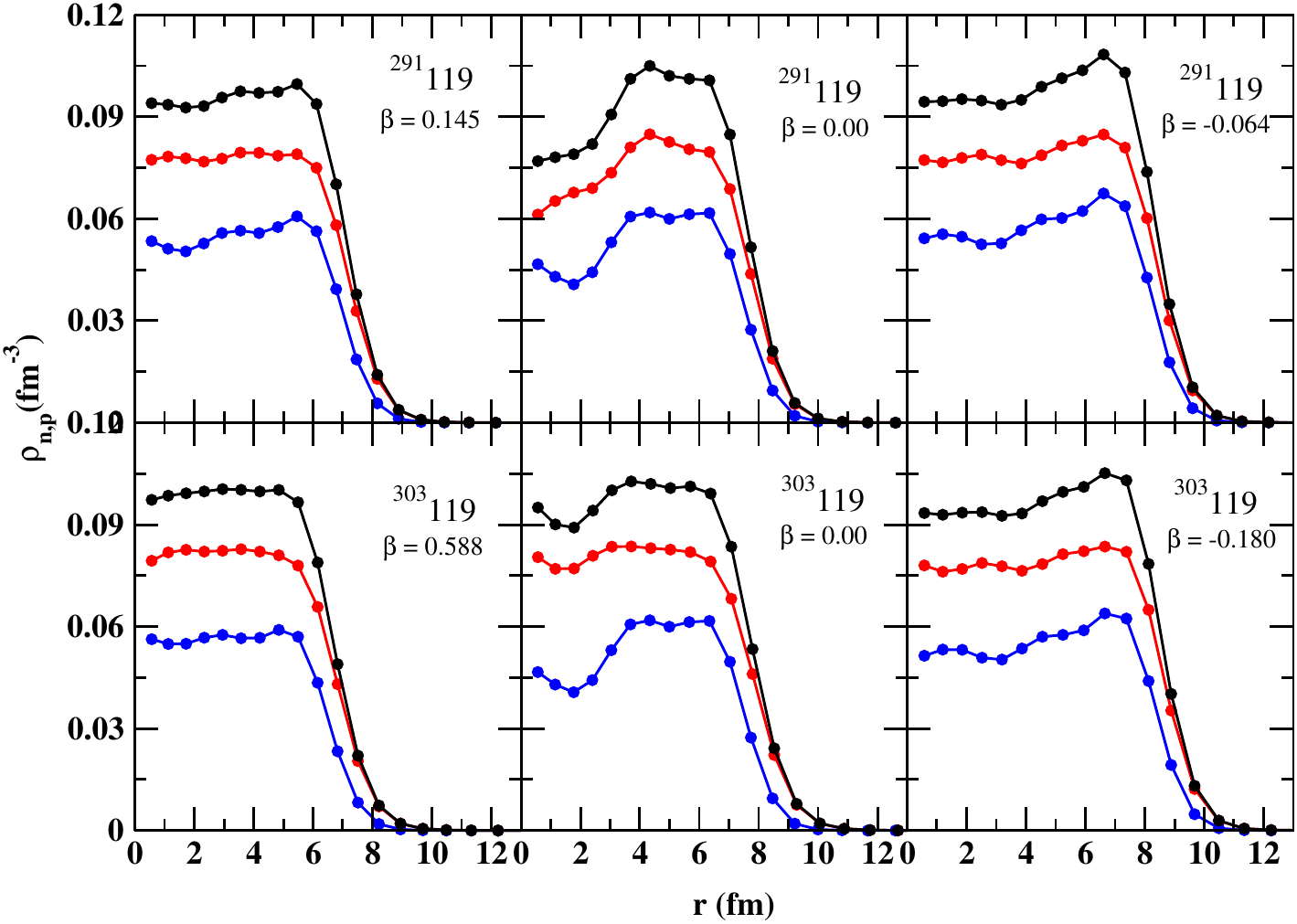}
	}
	\caption{(color online) Total, neutron and proton density distribution as a 
		function of radial parameter for three possible shape configurations. 
		Lines with black, red and blue colors represent 
		the total, neutron and proton density profile respectively.}
	\label{den1}
\end{figure}

\begin{figure}
\centering
	\resizebox{0.70\textwidth}{!}{%
		\includegraphics{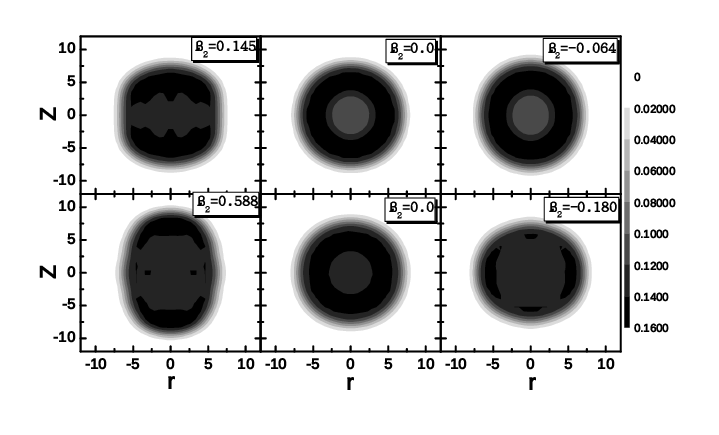}
	}
	\caption{ Two dimensional total matter density contours 
		of $^{291}119$ and $^{303}119$ for three different shape configurations. 
		Density profile of $^{291}119$ are seen in upper panel of the figure while 
		the density distribution of $^{303}119$ are represented in lower panel.}
	\label{den2}
\end{figure}

\begin{table*}
	\caption{Binding energy (BE), quadrupole deformation parameter ($\beta_2$) and radii 
		for $Z=119$ isotopic chain within three possible shape configurations.}
	\renewcommand{\tabcolsep}{0.19cm}
	\renewcommand{\arraystretch}{1.0}
	\footnotesize
	\begin{tabular}{c|ccc|ccc|ccc|ccc|cc}
		\hline\hline
		Nuclei&\multicolumn{3}{c|}{BE}&\multicolumn{3}{c|}{$\beta_2$}&\multicolumn{3}{c|}{$r_c$}&\multicolumn{3}{c|}{$r_t$}&\multicolumn{2}{c}{FRDM}\\
		\cline{2-4} \cline{5-7} \cline{8-10} \cline{11-13} \cline{14-15}
		&prol.&sph.&obl.&prol.&sph.&obl.&prol.&sph.&obl.&prol.&sph.&obl.&BE&$\beta_2$\\
		\hline
		$^{284}	119$&	2004.6& 2002.8&	      &	0.217& -0.033&	      &	6.298&	6.238&	     &  6.324&	6.263&	     &  1997.6&	0.072\\
		$^{285}	119$&	2012.6&	2011.9&	      & 0.203& -0.068&	      & 6.298&	6.248&	     &  6.326&	6.276&	     &  2006.6&	0.080\\
		$^{286}	119$&	2020.4&	2020.0&	      &	0.194&	0.045&	      &	6.299&	6.252&	     &	6.332&	6.283&	     &	2014.5&	0.080\\
		$^{287}	119$&	2028.2&	2028.4&	      &	0.187&	0.039&	      &	6.301&	6.257&	     &	6.338&	6.292&	     &	2023.6&	-0.104\\
		$^{288}	119$&	2035.7&	2036.7&	      &	0.181&	0.018&	      &	6.304&	6.262&	     &	6.345&	6.301&	     &	2031.1&	-0.104\\
		$^{289}	119$&	2042.9&	2044.9&	      &	0.175&	-0.002&	      &	6.307&	6.268&	     &	6.352&	6.310&	     &	2039.5&	0.089\\
		$^{290}	119$&	2050.1&	2052.9&	      &	0.168&	-0.001&	      &	6.309&	6.274&	     &	6.358&	6.319&	     &	2046.9&	0.089\\
		$^{291}	119$&	2057.0&	2060.6&	      &	0.145&	-0.001&	      &	6.301&	6.279&	     &	6.355&	6.328&	     &	2055.2&	0.081\\
		$^{292}	119$&	2063.8&	2067.8&	2066.2&	0.163&	-0.001&	-0.165&	6.317&	6.283&	6.312&	6.375&	6.337&	6.366&	2062.2&	0.089\\
		$^{293}	119$&	2069.6&	2074.6&	2073.6&	0.240&	-0.001&	-0.172&	6.370&	6.286&	6.322&	6.424&	6.345&	6.378&	2070.4&	0.081\\
		$^{294}	119$&	2076.4&	2081.1&	2080.4&	0.243&	-0.001&	-0.178&	6.379&	6.290&	6.330&	6.436&	6.353&	6.390&	2077.2&	0.081\\
		$^{295}	119$&	2083.0&	2087.6&	2087.2&	0.245&	-0.002&	-0.222&	6.386&	6.293&	6.364&	6.446&	6.361&	6.423&	2084.7&	0.072\\
		$^{296}	119$&	2089.5&	2094.0&	2094.2&	0.246&	-0.001&	-0.237&	6.393&	6.296&	6.381&	6.457&	6.369&	6.443&	2090.9&	-0.096\\
		$^{297}	119$&	2096.0&	2100.4&	2101.0&	0.247&	-0.001&	-0.244&	6.400&	6.299&	6.392&	6.467&	6.377&	6.457&	2098.6&	-0.079\\
		$^{298}	119$&	2110.2&	2106.8&	2107.5&	0.555&	-0.001&	-0.251&	6.609&	6.302&	6.403&	6.669&	6.385&	6.471&	2105.0&	-0.044\\
		$^{299}	119$&	2116.7&	2113.1&	2113.7&	0.559&	-0.001&	-0.259&	6.619&	6.305&	6.415&	6.683&	6.393&	6.486&	2112.0&	-0.018\\
		$^{300}	119$&	2122.9&	2119.4&	2119.8&	0.565&	0.000&	-0.266&	6.630&	6.308&	6.427&	6.697&	6.401&	6.501&	2117.8&	-0.008\\
		$^{301}	119$&	2129.1&	2125.5&	2125.8&	0.578&	0.000&	-0.273&	6.647&	6.311&	6.438&	6.718&	6.410&	6.515&	2124.3&	0.000\\
		$^{302}	119$&	2135.3&	2131.5&	2130.2&	0.586&	0.000&	-0.180&	6.661&	6.315&	6.378&	6.736&	6.418&	6.467&	2129.7&	0.000\\
		$^{303}	119$&	2141.2&	2137.1&	2136.1&	0.588&	0.000&	-0.180&	6.670&	6.320&	6.384&	6.748&	6.427&	6.477&	2135.9&	0.000\\
		$^{304}	119$&	2147.1&	2142.5&	2141.8&	0.589&	0.000&	-0.183&	6.678&	6.326&	6.392&	6.758&	6.437&	6.488&	2140.3&	0.000\\
		$^{305}	119$&	2152.8&	2147.6&	2147.3&	0.589&	0.000&	-0.188&	6.685&	6.333&	6.401&	6.769&	6.448&	6.500&	2146.5&	0.000\\
		$^{306}	119$&	2158.3&	2152.5&	2152.6&	0.592&	0.000&	-0.194&	6.694&	6.342&	6.412&	6.781&	6.459&	6.513&	2150.6&	0.001\\
		$^{307}	119$&	2163.6&	2157.4&	2158.0&	0.598&	-0.001&	-0.201&	6.706&	6.351&	6.422&	6.796&	6.470&	6.526&	2156.4&	0.000\\
		$^{308}	119$&	2168.8&	2162.3&	2163.3&	0.606&	0.012&	-0.208&	6.721&	6.360&	6.433&	6.813&	6.481&	6.540&	2160.6&	0.001\\
		$^{309}	119$&	2174.0&	2167.4&	2168.5&	0.613&	0.030&	-0.214&	6.734&	6.370&	6.443&	6.830&	6.493&	6.553&	2166.1&	0.001\\
		$^{310}	119$&	2178.9&	2172.4&	2173.6&	0.619&	0.039&	-0.218&	6.747&	6.380&	6.452&	6.846&	6.505&	6.566&	2170.4&	0.002\\
		$^{311}	119$&	2183.9&	2177.4&	2178.6&	0.627&	0.043&	-0.223&	6.760&	6.390&	6.462&	6.863&	6.517&	6.578&	2176.0&	0.003\\
		$^{312}	119$&	2190.7&	2182.4&	2183.5&	0.741&	0.044&	-0.227&	6.882&	6.399&	6.506&	6.981&	6.528&	6.591&	2180.1&	0.004\\
		$^{313}	119$&	2196.0&	2187.2&	2188.2&	0.743&	0.042&	-0.233&	6.892&	6.408&	6.483&	6.993&	6.539&	6.605&	2185.7&	0.004\\
		$^{314}	119$&	2197.2&	2192.0&	2193.0&	0.569&	0.035&	-0.240&	6.723&	6.416&	6.494&	6.838&	6.549&	6.619&	2196.3&	0.531\\
		$^{315}	119$&	2200.6&	2196.8&	2197.8&	0.458&	0.000&	-0.247&	6.637&	6.425&	6.506&	6.757&	6.559&	6.633&	2202.0&	0.541\\
		$^{316}	119$&	2205.6&	2201.6&	2202.4&	0.450&	0.000&	-0.253&	6.638&	6.434&	6.517&	6.761&	6.570&	6.648&	2206.4&	0.542\\
		$^{317}	119$&	2210.3&	2206.3&	2207.1&	0.445&	0.000&	-0.259&	6.640&	6.442&	6.527&	6.768&	6.580&	6.661&	2208.5&	0.360\\
		$^{318}	119$&	2214.9&	2210.6&	2211.7&	0.429&	0.000&	-0.264&	6.634&	6.449&	6.537&	6.767&	6.591&	6.674&	2212.5&	0.331\\
		$^{319}	119$&	2219.3&	2214.5&	2216.3&	0.386&	0.000&	-0.269&	6.606&	6.456&	6.547&	6.746&	6.601&	6.688&	2217.9&	0.331\\
		$^{320}	119$&	2224.0&	2218.2&	2220.7&	0.373&	0.000&	-0.274&	6.603&	6.462&	6.556&	6.747&	6.610&	6.701&	2222.0&	0.331\\
		$^{321}	119$&	2228.6&	2221.7&	2225.3&	0.360&	0.001&	-0.279&	6.600&	6.468&	6.565&	6.748&	6.620&	6.714&	2227.0&	0.331\\
		$^{322}	119$&	2233.2&	      &	2229.6&	0.336&	     &	-0.284&	6.589&	     &	6.573&	6.739&	     &	6.726&	2230.9&	0.322\\
		$^{323}	119$&	2237.7&	      &	2231.9&	0.328&	     &	-0.191&	6.591&	     &	6.526&	6.744&	     &	6.682&	2235.6&	0.322\\
		$^{324}	119$&	2242.2&	      &	2236.1&	0.320&	     &	-0.189&	6.592&	     &	6.532&	6.749&	     &	6.691&	2239.1&	0.322\\
		$^{325}	119$&	2246.7&	      &	2240.4&	0.313&	     &	-0.189&	6.594&	     &	6.538&	6.755&	     &	6.701&	2243.5&	0.322\\
		$^{326}	119$&	2250.9&	      &	2244.4&	0.307&	     &	-0.191&	6.596&	     &	6.545&	6.761&	     &	6.711&	2247.0&	0.312\\
		$^{327}	119$&	2255.1&	      &	2248.5&	0.300&	     &	-0.195&	6.597&	     &	6.552&	6.766&	     &	6.722&	2251.3&	0.313\\
		$^{328}	119$&	2259.2&	      &	2252.4&	0.292&	     &	-0.199&	6.596&	     &	6.559&	6.770&	     &	6.733&	2254.7&	0.314\\
		$^{329}	119$&	2263.3&	      &	2256.2&	0.283&	     &	-0.204&	6.596&	     &	6.566&	6.774&	     &	6.745&	2259.0&	0.325\\
		$^{330}	119$&	2267.4&	      &	2260.0&	0.275&	     &	-0.210&	6.596&	     &	6.574&	6.779&	     &	6.757&	2264.1&	0.405\\
		$^{331}	119$&	2271.4&	      &	2267.4&	0.269&	     &	-0.386&	6.597&	     &	6.725&	6.785&	     &	6.904&	2268.3&	0.394\\
		$^{332}	119$&	2275.4&	      &	2271.8&	0.267&	     &	-0.415&	6.602&	     &	6.760&	6.793&	     &	6.943&	2271.4&	0.394\\
		$^{333}	119$&	2279.2&	      &	2275.5&	0.267&	     &	-0.405&	6.607&	     &	6.755&	6.803&	     &	6.942&	2273.5&	0.116\\
		$^{334}	119$&	2283.0&	      &	2279.1&	0.268&	     &	-0.391&	6.613&	     &	6.748&	6.813&	     &	6.938&	2276.7&	0.117\\
		$^{335}	119$&	2286.7&	      &	2282.9&	0.268&	     &	-0.383&	6.619&	     &	6.745&	6.823&	     &	6.938&	2280.7&	0.108\\
		$^{336}	119$&	2290.5&	      &	2286.7&	0.269&	     &	-0.378&	6.625&	     &	6.747&	6.832&	     &	6.943&	2283.8&	0.108\\
		$^{337}	119$&	2294.1&	      &	2290.3&	0.269&	     &	-0.377&	6.630&	     &	6.751&	6.842&	     &	6.950&	2288.0&	0.108\\
		$^{338}	119$&	2297.8&	      &	2293.8&	0.268&	     &	-0.377&	6.636&	     &	6.757&	6.851&	     &	6.960&	2291.0&	0.108\\
		$^{339}	119$&	2301.3&	      &	2297.1&	0.265&	     &	-0.378&	6.641&	     &	6.764&	6.860&	     &	6.970&	2295.0&	0.108\\
		$^{340}	119$&	2304.8&	      &	2294.7&	0.261&	     &	-0.193&	6.647&	     &	6.613&	6.868&	     &	6.840&	&	\\
		$^{341}	119$&	2308.2&	      &	2298.0&	0.258&	     &	-0.188&	6.652&	     &	6.616&	6.876&	     &	6.847&	&	\\
		$^{342}	119$&	2311.2&	      &	2301.2&	0.255&	     &	-0.185&	6.658&	     &	6.620&	6.885&	     &	6.854&	&	\\
		$^{343}	119$&	2314.2&	      &	2304.3&	0.253&	     &	-0.183&	6.663&	     &	6.625&	6.894&	     &	6.862&	&	\\
		$^{344}	119$&	2317.0&	      &	2307.4&	0.251&	     &	-0.180&	6.669&	     &	6.629&	6.904&	     &	6.870&	&	\\
		\hline\hline
		\label{tab1}
	\end{tabular}
\end{table*}
\begin{table*}
	\caption{Table 1 is continued....}
	\renewcommand{\tabcolsep}{0.19cm}
	\renewcommand{\arraystretch}{1.0}
	\footnotesize
	\begin{tabular}{c|ccc|ccc|ccc|ccc|cc}
		\hline\hline
		Nuclei&\multicolumn{3}{c|}{BE}&\multicolumn{3}{c|}{$\beta_2$}&\multicolumn{3}{c|}{$r_c$}&\multicolumn{3}{c|}{$r_t$}&\multicolumn{2}{c}{FRDM}\\
		\cline{2-4} \cline{5-7} \cline{8-10} \cline{11-13} \cline{14-15}
		&prol.&sph.&obl.&prol.&sph.&obl.&prol.&sph.&obl.&prol.&sph.&obl.&BE&$\beta_2$\\
		\hline
		$^{345}	119$&	2319.8&	      &	2310.4&	0.249&	     &	-0.178&	6.674&	     &	6.634&	6.913&	     &	6.878&	&	\\
		$^{346}	119$&	2322.6&	      &	2313.3&	0.247&	     &	-0.176&	6.679&	     &	6.639&	6.922&	     &	6.886&	&	\\
		$^{347}	119$&	2325.3&	      &	2316.4&	0.245&	     &	-0.174&	6.683&	     &	6.643&	6.931&	     &	6.894&	&	\\
		$^{348}	119$&	2328.0&	      &	2319.3&	0.243&	     &	-0.172&	6.687&	     &	6.648&	6.940&	     &	6.902&	&	\\
		$^{349}	119$&	2330.7&	      &	2322.2&	0.241&	     &	-0.169&	6.691&	     &	6.653&	6.949&	     &	6.910&	&	\\
		$^{350}	119$&	2333.4&	      &	2325.2&	0.239&	     &	-0.166&	6.695&	     &	6.658&	6.958&	     &	6.918&	&	\\
		$^{351}	119$&	2335.9&	      &	2328.1&	0.237&	     &	-0.162&	6.699&	     &	6.663&	6.968&	     &	6.926&	&	\\
		$^{352}	119$&	2338.0&	      &	2331.0&	0.234&	     &	-0.159&	6.704&	     &	6.667&	6.977&	     &	6.934&	&	\\
		$^{353}	119$&	2339.7&	      &	2333.9&	0.228&	     &	-0.156&	6.708&	     &	6.672&	6.984&	     &	6.942&	&	\\
		$^{354}	119$&	2341.6&	      &	2336.8&	0.221&	     &	-0.155&	6.711&	     &	6.677&	6.991&	     &	6.951&	&	\\
		$^{355}	119$&	2343.4&	      &	2339.7&	0.214&	     &	-0.155&	6.714&	     &	6.682&	6.998&	     &	6.960&	&	\\
		$^{356}	119$&	2345.3&	      &	2342.5&	0.207&	     &	-0.155&	6.717&	     &	6.687&	7.005&	     &	6.969&	&	\\
		$^{357}	119$&	2347.2&	      &	2345.2&	0.194&	     &	-0.156&	6.720&	     &	6.692&	7.009&	     &	6.979&	&	\\
		$^{358}	119$&	2349.2&	      &	2347.6&	0.187&	     &	-0.181&	6.723&	     &	6.712&	7.016&	     &	7.001&	&	\\
		$^{359}	119$&	2351.1&	      &	2350.1&	0.181&	     &	-0.188&	6.725&	     &	6.723&	7.023&	     &	7.015&	&	\\
		$^{360}	119$&	2352.9&	      &	2352.2&	0.176&	     &	-0.185&	6.725&	     &	6.727&	7.031&	     &	7.023&	&	\\
		$^{361}	119$&	2354.8&	      &	2354.3&	0.171&	     &	-0.179&	6.726&	     &	6.729&	7.039&	     &	7.030&	&	\\
		$^{362}	119$&	2356.8&	      &	2356.2&	0.166&	     &	-0.173&	6.728&	     &	6.730&	7.048&	     &	7.036&	&	\\
		$^{363}	119$&	2358.6&	2357.8&	2358.2&	0.161&	0.091&	-0.167&	6.730&	6.701&	6.731&	7.056&	     &	7.043&	&	\\
		$^{364}	119$&	2360.3&	2359.9&	2360.3&	0.153&	0.083&	-0.160&	6.730&	6.703&	6.732&	7.062&	7.028&	7.050&	&	\\
		$^{365}	119$&	2361.8&	2362.1&	2362.4&	0.143&	0.064&	-0.154&	6.729&	6.705&	6.733&	7.066&	7.034&	7.057&	&	\\
		$^{366}	119$&	2363.1&	2364.3&	2364.5&	0.162&	0.053&	-0.154&	6.743&	6.708&	6.738&	7.089 &	7.042&	7.067&	&	\\
		$^{367}	119$&	2364.3&	2366.4&	2366.4&	0.168&	0.047&	-0.154&	6.752&	6.710&	6.743&	7.106&	7.051&	7.077&	&	\\
		$^{368}	119$&	2365.7&	2368.5&	2367.9&	0.177&	0.036&	-0.150&	6.772&	6.711&	6.746&	7.111&	7.059&	7.086&	&	\\
		$^{369}	119$&	2366.2&	2370.5&	2369.3&	0.195&	0.009&	-0.138&	6.793&	6.712&	6.746&	7.131&	7.067&	7.092&	&	\\
		$^{370}	119$&	2367.7&	2372.7&	2370.4&	0.203&	0.002&	-0.185&	6.806&	6.714&	6.782&	7.145&	7.076&	7.122&	&	\\
		$^{371}	119$&	2369.1&	2374.7&	2372.0&	0.206&	0.001&	-0.201&	6.815&	6.716&	6.801&	7.156&	7.086&	7.140&	&	\\
		$^{372}	119$&	2370.2&	2376.5&	2373.3&	0.209&	0.001&	-0.207&	6.824&	6.718&	6.811&	7.168&	7.096&	7.154&	&	\\
		$^{373}	119$&	2371.3&	2378.2&	2374.5&	0.215&	0.001&	-0.215&	6.834&	6.719&	6.823&	7.182&	7.106&	7.169&	&	\\
		$^{374}	119$&	2372.4&	2379.9&	2375.6&	0.224&	0.001&	-0.228&	6.842&	6.720&	6.841&	7.195&	7.116&	7.187&	&	\\
		$^{375}	119$&	2373.5&	2381.5&	2377.0&	0.236&	0.001&	-0.238&	6.849&	6.721&	6.857&	7.209&	7.126&	7.204&	&	\\
		\hline\hline
		\label{tab2}
	\end{tabular}
\end{table*}

\subsection{ Neutron-separation energy}
Separation energy is the first prime signature to identify 
the magic behaviour of the nuclei. 
The magic numbers in nuclei are characterized 
by large shell gaps in their single-particle energy levels. 
Large shell gap means the nucleons occupying the lower energy level have 
comparatively large value of energy than those nucleons occupying the 
higher energy levels. 
This large energy difference between two consecutive energy levels 
can be observed from the sudden fall of neutron separation energy 
which attribute the extra stability to a particular nucleus 
having certain numbers of nucleons and that's why closed shell 
nuclei are more bound than their nearby ones.
Moreover, two-neutron separation energy is more significant than one neutron 
due to its takes care of even-odd staggering and it, therefore, 
manifests the magicity more clearly.
One and two-neutron separation energy is calculated by the 
difference in binding energies of two isotopes using the relations
\begin{eqnarray}
S_{n}(N,Z) &= BE (N,Z) - BE(N-1,Z),\nonumber \\
S_{2n}(N,Z)&= BE (N,Z) - BE (N-2,Z).
\end{eqnarray}                                                                   
One- and two-neutron separation energy for the considered isotopic 
series of the nuclei $^{284-375}$119 are plotted in lower panel of Fig. 1.
No sudden fall of the separation energies is noticed in present analysis which 
indicates that as such no neutron magic behaviour 
within this force parameter is exhibited.
\subsection{Shape Coexistence}
The shape of a nucleus is one of the fundamental properties along 
with its mass and radius. 
It is the result of the interplay between  macroscopic liquid-drop
like properties of the nuclear matter and microscopic shell effects. 
In some areas of the nuclear chart, the shape is seen to be very sensitive to 
structural effect and may change from one nucleus to its neighbour. 
These changes are caused by the rearrangement of the orbital configuration 
of the nucleons or by the dynamic response of the nucleus to rotation. 
However, there might arise a situation where we may witness that 
configurations corresponding to different shapes may coexist at similar 
energies or by a very little difference.
The small binding energy difference between two shape configuration makes the 
structure more complex and the study of such nuclei enrich our 
understanding of the oscillations of nuclei occurring between two 
or three existing shapes.
This leads isomers can appear in superheavy region.
The phenomenon of shape coexistence is ubiquitous as it has been 
observed throughout the nuclear landscape starting from light 
nuclei~\cite{morinaga1956} to the regions of heavy 
nuclei~\cite{heyde1984,andreyev2000} and ofcourse in superheavy 
region~\cite{ren2002,zheng2008,zhao2015,cwiok2000,ren2001}. 
No case of shape-coexistence is observed in considered isotopic chain of $Z=119$. 
However, shape-coexistence can be a common phenomenon in superheavy nuclei and 
thus it is interesting to study it by future experiments.
Here, we noticed a very little energy difference around $\le$ 1 MeV 
in first and second intrinsic excited states in some of the nuclides. 
For example, in $^{298-318}119$ nuclides the excited energy differed by 
the amount of $\le$ 1 MeV within spherical and oblate configurations, 
whereas prolate suggested to be ground state. 
\subsection{Density profile}
In general, the neutron excess becomes larger with increasing the mass 
number and ofcourse it is quite natural in case of superheavy nuclei 
providing the largest neutron excesses.
However, these nuclei also have large number of protons and 
therefore huge Coulomb repulsion exist there that pushes the proton 
to larger radii and as a result change the proton density distribution.  
In this view, neutron and proton density distributions are 
considered to be great source of potential providing 
fundamental information on nuclear structure and quite useful 
to identify the special kinds of features of nuclei 
such as Bubble, Halo/Skin and cluster structures. 
Such features are observed in light to superheavy 
nuclei~\cite{whee,wilson1946,decharge2003,grasso2009,singh2013,sharma2006}.
In the search of such exotic structures, we have made the plot for 
total, neutron and proton density profile for predicting neutron shell closure 
nuclei $^{291}119$ and $^{303}119$ within this framework as shown in Fig. 3.
The spherical configuration of these two nuclides 
show the depletion of central part of neutron, proton and 
total matter (neutron plus proton) density.
At prolate configuration the density dies at $r=8$ $fm$ while it 
reaches to 10 $fm$ in oblate and spherical configurations.
This distribution signals prolate as a ground state of these nuclides.
Moreover, spherical structure of these two nuclides indicates 
a special kind of proton distribution.
In which, the centre is little bulgy and a considerably depletion 
afterward but again a big hump and further distribution tends to 
zero at the end of the surface following a decreasing pattern.
To reveal such anomalous behaviour of nucleon distribution and to visualize the 
arrangement of nucleons more clearly inside the nuclei, we make two-dimensional 
contour plots for $^{291}119$ and $^{303}119$ within three different 
shape configurations as given in Fig. 4. 
The full black contour refers to maximum density and full white ones
to zero density region.
Figure 4 reflects that the hollow region at the centre is spread 
over the radius of $1-3$ $fm$ in spherical configuration. 
A considerable depopulation is revealed in spherical shape which 
may supposed to be semi-bubble type structure.
It is also apparent that the region from $3-6$ $fm$ of 
total matter density distribution in both the nuclei is 
highly dense and formed a thick ring type structure. 
It can be interpreted as a some how hollow central part is surrounded by a 
thick sheath of nucleons (high density) and formed a thick ring type 
structure in prolate shaped.
For both the nuclei, in prolate and oblate configurations, the 
matter distribution is not uniform and bunches of nucleons far 
from the centre is seen. 
These bunches may be the cluster of nucleons or alpha particles. 
Some spindle type structure is also noticed in prolate configuration 
having flaps/bulges shapes.
In general, cluster, semi-bubble as well as thick ring type structure is 
seen.

\begin{table*}
	\caption{Decay energies (in MeV) and half-lives of 
		$\alpha$, $\beta$ and spontaneous fission for $Z=119$ isotopic chain and 
		prediction of mode of decays is given.}
	\renewcommand{\tabcolsep}{0.03cm}
	\renewcommand{\arraystretch}{1.0}
	\footnotesize\footnotesize 
	\begin{tabular}{ccccccccccccccc}
		\hline\hline
		Nuclei&$Q_\alpha^{RMF}$&$Q_\alpha^{FRDM}$&\multicolumn{6}{c}{$\log$($T_{1/2}^\alpha$)}&$\log$($T_{1/2}^{SF}$)&$Q_\beta^{RMF}$&$Q_\beta^{FRDM}$&$\log$($T_{1/2}^\beta)$&$T_{1/2}^\beta$(sec)&Mode of \\
		\cline{4-9} 
		&&&VSS&Brown&Royer&GLDM&Ni et. al.&FRDM&Ren-Xu&&&{Fiset-Nix}&FRDM&decay\\
		\hline
		$^{284}	119$&	14.00	&13.02	&-6.29	&-6.98 &-6.86	&-6.64&	-7.29&	-4.42&	-6.33		&14.29&	10.35	&0.03&	3.73&$\alpha$\\
		$^{285}	119$&	13.89	&13.79	&-6.43	&-6.79 &-6.67	&-6.81&	-7.66&	-6.25&	-1.77		&14.41&	8.80	&0.05&	3.63&$\alpha$\\
		$^{286}	119$&	13.83	&13.74	&-5.97	&-6.69 &-6.57	&-6.33&	-7.00&	-5.81&	1.95		&14.47&	9.69	&0.03&	5.75&$\alpha$\\
		$^{287}	119$&	13.79	&13.45	&-6.26	&-6.63 &-6.53	&-6.67&	-7.51&	-5.60&	4.84		&14.50&	7.97	&0.05&	4.45&$\alpha$\\
		$^{288}	119$&	13.99	&13.38	&-6.27	&-6.95 &-6.90	&-6.69&	-7.26&	-5.14&	6.90		&14.30&	8.87	&0.03&	5.82&$\alpha$\\
		$^{289}	119$&	14.12	&13.35	&-6.84	&-7.15 &-7.15	&-7.30&	-8.01&	-5.79&	8.15		&7.92&	7.55	&0.22&	12.84&$\alpha$\\
		$^{290}	119$&	14.22	&13.36	&-6.68	&-7.31 &-7.35	&-7.16&	-7.61&	-5.09&	8.59		&7.67&	8.21	&0.06&	21.97&$\alpha$\\
		$^{291}	119$&	14.38	&13.20	&-7.30	&-7.56 &-7.64	&-7.80&	-8.40&	-5.13&	8.24		&7.43&	6.76	&0.38&	41.09&$\alpha$\\
		$^{292}	119$&	14.45	&13.17	&-7.08 	&-7.66 &-7.78	&-7.62&	-7.95&	-4.71&	7.10		&6.53&	7.71	&0.45&	27.00&$\alpha$\\
		$^{293}	119$&	15.32	&12.88	&-8.86	&-8.92 &-9.22	&-9.39&	-9.73&	-4.47&	5.17		&6.30&	6.03	&0.78&	73.84&$\alpha$\\
		$^{294}	119$&	15.13	&12.80	&-8.22	&-8.66 &-8.95	&-8.87&	-8.93&	-3.97&	2.47		&5.98&	6.85	&0.66&	80.88&$\alpha$\\
		$^{295}	119$&	16.18	&12.88  &-10.16	&-10.06&-10.56	&-10.73&-10.85&	-4.49&	-0.99		&4.81&	5.45	&1.42&	$>$100&$\alpha$\\
		$^{296}	119$&	16.02	&13.08	&-9.58	&-9.86 &-10.34	&-10.36&-10.10&	-4.54&	-5.21		&5.41&	6.77	&0.91&	28.29&$\alpha$\\
		$^{297}	119$&	16.03	&12.74	&-9.95	&-9.88 &-10.38	&-10.56&-10.67&	-4.19&	-10.19		&5.10&	5.19	&1.29&	71.00&SF\\
		$^{298}	119$&	11.48	&12.50	&-1.01	&-2.33 &-1.84	&-1.25&	-2.76&	-3.34&	-15.90		&5.01&	5.68	&1.09&	$>$100&SF\\
		$^{299}	119$&	11.37	&12.80	&-1.09	&-2.10 &-1.60	&-1.72&	-3.09&	-4.32&	-22.35		&3.38&	4.18	&2.25&	$>$100&SF\\
		$^{300}	119$&	9.96	&13.15  & 3.11	&1.28  & 2.22	&3.11&	0.76&	-4.67&	-29.53		&9.70&	4.92	&-0.51&	$>$100&SF\\
		$^{301}	119$&	11.04	&13.27	&-0.25	&-1.30 &-0.80	&-0.92&	-2.37&	-5.25&	-37.43		&4.38&	3.55	&1.66&	$>$100&SF\\
		$^{302}	119$&	11.09	&13.38	&-0.04	&-1.48 &-0.95	&-0.29&	-1.93&	-5.13&	-46.04		&4.07&	4.34	&1.58&	$>$100&SF\\
		$^{303}	119$&	11.24	&13.38	&-0.76	&-1.82 &-1.34	&-1.47&	-2.81&	-5.46&	-55.36		&3.72&	2.16	&2.04&	$>$100&SF\\
		$^{304}	119$&	11.65	&14.14	&-1.42	&-2.70 &-2.36	&-1.79&	-3.11&	-6.55&	-65.37		&3.27&	3.99	&2.09&	$>$100&SF\\
		$^{305}	119$&	11.95	&13.84	&-2.47	&-3.31 &-3.07	&-3.21&	-4.27&	-6.33&	-76.08		&2.83&	1.59	&2.66&	$>$100&SF\\
		$^{306}	119$&	6.44	&13.97	&17.69	&14.09 &16.65	&18.57&	13.25&	-6.23&	-87.48		&8.25&	3.28	&-0.10&	$>$100&SF\\
		$^{307}	119$&	5.88	&13.81	&20.81	&17.12 &20.07	&20.06&	15.65&	-6.29&	-99.55		&8.43&	1.27	&0.09&	$>$100&SF\\
		$^{308}	119$&	5.31	&13.43	&25.26	&20.73 &24.14	&26.61&	19.72&	-5.23&	-112.30		&1.61&	2.64	&3.60&	$>$100&SF\\
		$^{309}	119$&	9.03	&13.31	&5.76 	&3.91  &5.05	&4.95&	2.77&	-3.35&	-125.71		&1.36&	0.96	&4.18&	$>$100&SF\\
		$^{310}	119$&	8.88	&12.76	&6.63 	&4.38  &5.57	&6.70&	3.78&	-3.88&	-139.79		&1.08&	1.99	&4.35&	$>$100&SF\\
		$^{311}	119$&	8.91	&12.49	&6.21 	&4.30  &5.46	&5.37&	3.16&	-3.65&	-154.51		&0.81&	0.51	&5.08&	$>$100&SF\\
		$^{312}	119$&	6.81	&12.11	&15.63	&12.27 &14.49	&16.27&	11.48&	-2.48&	-169.88		&1.02&	$\pm$	&4.46&	$\pm$&SF\\
		$^{313}	119$&	6.30	&13.25	&18.17	&14.80 &17.34	&17.31&	13.40&	-5.22&	-185.90		&0.69&	6.28	&5.33&	14.29&SF\\
		$^{314}	119$&	8.28	&4.66	&8.91 	&6.37  &7.76	&9.06&	5.73&	$>$20&	-202.54		&-3.42&	1.30	&2.00&	$>$100&SF\\
		$^{315}	119$&	9.34	&4.13	&4.70 	&2.98  &3.89	&3.79&	1.87&	$>$20&	-219.82		&-4.74&	$\pm$	&1.49&	stable&SF\\
		$^{316}	119$&	9.17	&5.14	&5.64 	&3.50  &4.47	&5.54&	2.93&	$>$20&	-237.72		&0.45&	1.27	&5.70&	$>$100&SF\\
		$^{317}	119$&	9.25	&8.14	&5.01 	&3.25  &4.17	&4.07&	2.14&	 9.11&	-256.23		&0.41&	3.77	&6.04&	1.14&SF\\
		$^{318}	119$&	9.33	&7.84	&5.08 	&3.01  &3.88	&4.91&	2.45&	10.74&	-275.36		&0.05&	$\pm$	&7.47&	$\pm$&SF\\
		$^{319}	119$&	9.40	&8.74	&4.51 	&2.81  &3.64	&3.54&	1.71&	 6.79&	-295.09		&-0.12&	0.57	&7.18&	$>$100&SF\\
		$^{320}	119$&	9.07	&8.68	&5.97 	&3.80  &4.74	&5.83&	3.22&	 7.36&	-315.42		&-0.15&	2.05	&6.77&	$>$100&SF\\
		$^{321}	119$&	8.71	&8.71	&6.91 	&4.92  &5.99	&5.90&	3.76&	 6.92&	-336.34		&-0.34&	1.19	&6.27&	$>$100&SF\\
		$^{322}	119$&	8.42	&8.72	&8.36 	&5.90  &7.09	&8.35&	5.26&	 7.21&	-357.85		&-0.50&	2.69	&5.56&	$>$100&SF\\
		$^{323}	119$&	8.17	&8.98	&9.01 	&6.76  &8.05	&7.97&	5.56&	 5.94&	-379.94		&-0.53&	1.89	&5.72&	$>$100&SF\\
		$^{324}	119$&	8.01	&9.00	&10.00	&7.33  &8.69	&10.06&	6.67&	 6.21&	-402.62		&-0.70&	3.36	&5.09&	71.06&SF\\
		$^{325}	119$&	7.80	&8.94	&10.57	&8.13  &9.57	&9.50&	6.89&	 6.09&	-425.86		&-0.88&	2.41	&4.96&	$>$100&SF\\
		$^{326}	119$&	7.87	&8.72	&10.61	&7.87  &9.26	&10.68&	7.18&	 7.21&	-449.67		&-1.12&	3.73	&4.31&	22.28&SF\\
		$^{327}	119$&	7.82	&8.55	&10.48	&8.05  &9.45	&9.38&	6.82&	 7.52&	-474.04		&-1.33&	2.82	&4.24&	17.84&SF\\
		$^{328}	119$&	7.80	&8.27	&10.88	&8.11  &9.50	&10.94&	7.42&	 8.95&	-498.97		&-1.47&	4.08	&3.81&	31.41&SF\\
		$^{329}	119$&	7.72	&8.08	&10.91	&8.43  &9.85	&9.77&	7.18&	 9.35&	-524.45		&-1.57&	3.16	&3.93&	54.58&SF\\
		$^{330}	119$&	7.53	&6.24	&12.11	&9.19  &10.69	&12.22&	8.47&	18.92&	-550.47		&-1.66&	2.57	&3.57&	$>$100&SF\\
		$^{331}	119$&	7.49	&6.03	&11.92	&9.32  &10.82	&10.75&	8.05&	19.87&	-577.04		&-1.80&	1.65	&3.66&	$>$100&SF\\
		$^{332}	119$&	7.45	&5.98	&12.48	&9.51  &11.02	&12.58&	8.79&	$>$20&	-604.14		&-2.01&	2.96	&3.19&	$>$100&SF\\
		
		\hline\hline
		\label{tab3}
	\end{tabular}
\end{table*}

\begin{table*}
	\caption{Table 3 is continued.....}
	\renewcommand{\tabcolsep}{0.07cm}
	\renewcommand{\arraystretch}{1.0}
	\footnotesize\footnotesize
	\begin{tabular}{ccccccccccccccc}
		\hline\hline
		Nuclei&$Q_\alpha^{RMF}$&$Q_\alpha^{FRDM}$&\multicolumn{6}{c}{$\log$($T_{1/2}^\alpha$)}&$\log$($T_{1/2}^{SF}$)&$Q_\beta^{RMF}$&$Q_\beta^{FRDM}$&$\log$($T_{1/2}^\beta)$&$T_{1/2}^\beta$(sec)&Mode of \\
		\cline{4-9} 
		&&&VSS&Brown&Royer&GLDM&Ni et. al.&FRDM&Ren-Xu&&&{Fiset-Nix}&FRDM&decay\\
		\hline
		$^{333}	119$&	7.44	&7.82	&12.16	&9.53  &11.03	&10.96&	8.26&	10.48&	-631.78		&-2.22&	5.96	&3.22&	0.08&SF\\
		$^{334}	119$&	7.33	&7.57	&13.05	&10.01 &11.56	&13.15&	9.27&	 11.9&	-659.94		&-2.41&	4.85	&2.80&	$>$100&SF\\
		$^{335}	119$&	7.17	&7.42	&13.46	&10.67 &12.29	&12.23&	9.37&	12.28&	-688.62		&-2.60&	3.61	&2.88&	$>$100&SF\\
		$^{336}	119$&	6.91	&7.16	&15.12	&11.83 &13.59	&15.33&	11.05&	13.86&	-717.83		&-2.79&	5.02	&2.49&	4.52&SF\\
		$^{337}	119$&	6.72	&6.76	&15.78	&12.70 &14.57	&14.51&	11.35&	15.58&	-747.54		&-2.97&	3.86	&2.60&	43.33&SF\\
		$^{338}	119$&	6.53	&6.54	&17.17	&13.62 &15.59	&17.48&	12.80&	17.11&	-777.76		&-3.15&	5.33	&2.22&	9.55&SF\\
		$^{339}	119$&	6.39	&6.36	&17.63	&14.33 &16.38	&16.34&	12.94&	17.82&	-808.49		&-3.38&	4.32	&2.30&	&SF\\
		$^{340}	119$&	6.32	&	&18.44	&14.74 &16.83	&18.81&	13.89&	&	-839.72		&-3.65&		&1.89&	&SF\\
		$^{341}	119$&	6.12	&	&19.29	&15.79 &18.00	&17.96&	14.35&	&	-871.44		&-3.99&		&1.93&	&SF\\
		$^{342}	119$&	6.09	&	&19.81	&15.95 &18.17	&20.24&	15.06&	&	-903.65		&-4.46&		&1.43&	&SF\\
		$^{343}	119$&	6.00	&	&20.05	&16.46 &18.73	&18.70&	15.01&	&	-936.35		&-4.80&		&1.50&	&SF\\
		$^{344}	119$&	5.89	&	&21.08	&17.06 &19.40	&21.57&	16.15&	&	-969.53		&-5.03&		&1.15&	&SF\\
		$^{345}	119$&	5.61	&	&22.72&18.80&	21.36	&21.34&	17.30&	&	-1003.18	&-5.25&		&1.29&	&SF\\
		$^{346}	119$&	5.30	&	&25.32&20.79&	23.60	&26.07&	19.78&	&	-1037.32	&-5.49&		&0.95&	&SF\\
		$^{347}	119$&	5.16	&	&26.12&21.79&	24.71	&24.71&	20.20&	&	-1071.92	&-5.68&		&1.11&	&SF\\
		$^{348}	119$&	4.90	&	&28.66&23.71&	26.88	&29.59&	22.63&	&	-1106.98	&-5.82&		&0.81&	&SF\\
		$^{349}	119$&	4.67	&	&30.35&25.50&	28.89	&28.91&	23.82&	&	-1142.51	&-5.93&		&1.01&	&SF\\
		$^{350}	119$&	4.48	&	&32.58&27.16&	30.76	&33.75&	25.99&	&	-1178.50	&-6.05&		&0.72&	&SF\\
		$^{351}	119$&	4.49	&	&32.13&27.06&	30.63	&30.66&	25.34&	&	-1214.94	&-6.29&		&0.87&	&SF\\
		$^{352}	119$&	5.00	&	&27.78&22.94&	25.95	&28.59&	21.88&	&	-1251.83	&-6.88&		&0.41&	&SF\\
		$^{353}	119$&	5.65	&	&22.40&18.52&	20.91	&20.89&	17.02&	&	-1289.16	&-7.24&		&0.53&	&SF\\
		$^{354}	119$&	5.69	&	&22.46&18.27& 	20.62	&22.88&	17.33&	&	-1326.94	&-7.32&		&0.27&	&SF\\
		$^{355}	119$&	5.56	&	&23.07&19.11&	21.56	&21.53&	17.60&	&	-1365.16	&-7.37&		&0.49&	&SF\\
		$^{356}	119$&	5.38	&	&24.70&20.24&	22.82	&25.25&	19.25&	&	-1403.81	&-7.45&		&0.23&	&SF\\
		$^{357}	119$&	5.19	&	&25.84&21.54&	24.28	&24.27&	19.96&	&	-1442.89	&-7.43&		&0.48&	&SF\\
		$^{358}	119$&	5.05	&	&27.36&22.57&	25.44	&28.05&	21.52&	&	-1482.40	&-7.48&		&0.22&	&SF\\
		$^{359}	119$&	4.86	&	&28.67&24.02&	27.07	&27.07&	22.38&	&	-1522.34	&-7.60&		&0.42&	&SF\\
		$^{360}	119$&	4.91	&	&28.58&23.64&	26.62	&29.32&	22.57&	&	-1562.69	&-7.69&		&0.15&	&SF\\
		$^{361}	119$&	4.86	&	&28.60&23.96&	26.96	&26.97&	22.32&	&	-1603.47	&-7.71&		&0.39&	&SF\\
		$^{362}	119$&	4.67	&	&30.69&25.50&	28.70	&31.55&	24.37&	&	-1644.65	&-7.73&		&0.14&	&SF\\
		$^{363}	119$&	4.61	&	&30.97&26.04&	29.30	&29.31&	24.35&	&	-1686.25	&-7.83&		&0.36&	&SF\\
		$^{364}	119$&	4.80	&	&29.47&24.43&	27.45	&30.22&	23.33&	&	-1728.25	&-10.22&	&-0.54&	&SF\\
		$^{365}	119$&	4.97	&	&27.65&23.13&	25.96	&25.96&	21.51&	&	-1770.65	&-8.13&		&0.27&	&SF\\
		$^{366}	119$&	4.11	&	&36.49&30.59&	34.41	&37.68&	29.34&	&	-1813.46	&-6.12&		&0.72&	&SF\\
		$^{367}	119$&	3.55	&	&43.17&36.76&	41.39	&41.47&	34.80&	&	-1856.66	&-5.67&		&1.14&	&SF\\
		$^{368}	119$&	6.37	&	&18.11&14.45&	16.07	&18.03&	13.60&	&	-1900.26	&-9.08&		&-0.24&	&SF\\
		$^{369}	119$&	6.63	&	&16.30&13.16&	14.60	&14.53&	11.79&	&	-1944.24	&-9.16&		&-0.02&	&SF\\
		$^{370}	119$&	7.02	&	&14.58&11.36&	12.53	&14.24&	10.59&	&	-1988.62	&-9.28&		&-0.29&	&SF\\
		$^{371}	119$&	3.48	&	&44.22&37.67&	42.38	&42.46&	35.69&	&	-2033.38	&-9.41&		&-0.08&	&SF\\
		$^{372}	119$&	3.39	&	&45.85&38.81&	43.65	&47.58&	37.35&	&	-2078.51	&-9.66&		&-0.39&	&SF\\
		$^{373}	119$&	3.27	&	&47.46&40.52&	45.57	&45.67&	38.46&	&	-2124.03	&-9.92&		&-0.21&	&SF\\
		$^{374}	119$&	3.07	&	&51.07&43.39&	48.81	&53.12&	41.81&	&	-2169.92	&-10.15&	&-0.51&	&SF\\
		$^{375}	119$&	2.99	&	&52.30&44.77&	50.37	&50.49&	42.61&	&	-2216.19	&-10.18&	&-0.27&	&SF\\
		\hline\hline
		\label{tab4}
	\end{tabular}
\end{table*}

\begin{table*}
	\caption{Same as Table III and IV but for the study of $\alpha-$decay chains of fission survival 
		nuclides (i.e.$^{284-296}119$) of the considered isotopic chain. Experimental data for $Q_\alpha$~\cite{expq1,expq2,hamilton2012}, 
		if available, is given in parentheses with asterisk.}
	\renewcommand{\tabcolsep}{0.15cm}
	\renewcommand{\arraystretch}{1.0}
	\footnotesize\footnotesize
	\begin{tabular}{ccccccccc}
		\hline\hline
		Nuclei&$Q_\alpha^{RMF}$&\multicolumn{5}{c}{$T_{1/2}^\alpha$}&{$T_{1/2}^{SF}$}&Mode of \\
		\cline{3-7} 
		&&VSS&Brown&Royer&GLDM&Ni et. al.&Ren-Xu&decay\\
		\hline
		$^{284}119$&	14.005& 0.502 $\times$  $10^{-06}$&  0.106 $\times$  $10^{-06}$&  0.139 $\times$  $10^{-06}$& 	0.227 $\times$  $10^{-06}$&0.518 $\times$  $10^{-07}$& 1.174&$\alpha$1\\	
		$^{280}$Ts&	13.257& 0.398 $\times$  $10^{-05}$&  0.615 $\times$  $10^{-06}$&  0.102 $\times$  $10^{-05}$& 	0.183 $\times$  $10^{-05}$&0.286 $\times$  $10^{-06}$&-3.328&SF\\	
		$^{276}$Mc&	12.353& 0.783 $\times$  $10^{-04}$&  0.836 $\times$  $10^{-05}$&  0.189 $\times$  $10^{-04}$& 	0.392 $\times$  $10^{-04}$&0.347 $\times$  $10^{-05}$&-6.446&SF\\	
		$^{272}$Nh&	12.032& 0.107 $\times$  $10^{-03}$&  0.109 $\times$  $10^{-04}$&  0.245 $\times$  $10^{-04}$& 	0.490 $\times$  $10^{-04}$&0.432 $\times$  $10^{-05}$&-8.257&SF\\	
		$^{268}$Rg&	11.714& 0.144 $\times$  $10^{-03}$&  0.142 $\times$  $10^{-04}$&  0.317 $\times$  $10^{-04}$& 	0.611 $\times$  $10^{-04}$&0.536 $\times$  $10^{-05}$&-8.839&SF\\	
		&&&&&&&&\\
		$^{285}119$&	13.891& 0.368 $\times$  $10^{-06}$&  0.160 $\times$  $10^{-06}$&  0.214 $\times$  $10^{-06}$& 	0.153 $\times$  $10^{-06}$& 0.219 $\times$  $10^{-07}$& 5.732&$\alpha$1\\	
		$^{281}$Ts&	13.396& 0.982 $\times$  $10^{-06}$&  0.358 $\times$  $10^{-06}$&  0.533 $\times$  $10^{-06}$& 	0.388 $\times$  $10^{-06}$&0.475 $\times$  $10^{-07}$& 0.977&$\alpha$2\\	
		$^{277}$Mc&	12.286& 0.494 $\times$  $10^{-04}$&  0.112 $\times$  $10^{-04}$&  0.250 $\times$  $10^{-04}$& 	0.189 $\times$  $10^{-04}$&0.129 $\times$  $10^{-05}$&-2.409&SF\\	
		$^{273}$Nh&	11.873& 0.108 $\times$  $10^{-03}$&  0.222 $\times$  $10^{-04}$&  0.519 $\times$  $10^{-04}$& 	0.399 $\times$  $10^{-04}$&0.240 $\times$  $10^{-05}$&-4.499&SF\\	
		$^{269}$Rg&	11.398& 0.335 $\times$  $10^{-03}$&  0.619 $\times$  $10^{-04}$&  0.155 $\times$  $10^{-03}$& 	0.122 $\times$  $10^{-03}$&0.609 $\times$  $10^{-05}$&-5.373&SF\\	
		&&&&&&&&\\
		$^{286}119$&	13.827& 0.106 $\times$  $10^{-05}$&  0.203 $\times$  $10^{-06}$&  0.269 $\times$  $10^{-06}$&   0.464 $\times$  $10^{-06}$&0.981 $\times$  $10^{-07}$& 9.451&$\alpha$1\\	
		$^{282}$Ts&	13.296& 0.335 $\times$  $10^{-05}$&  0.528 $\times$  $10^{-06}$&  0.792 $\times$  $10^{-06}$&   0.140 $\times$  $10^{-05}$&0.246 $\times$  $10^{-06}$& 4.491&$\alpha$2\\	
		$^{278}$Mc&	12.306& 0.984 $\times$  $10^{-04}$&  0.103 $\times$  $10^{-04}$&  0.218 $\times$  $10^{-04}$&   0.460 $\times$  $10^{-04}$& 0.422 $\times$  $10^{-05}$& 0.891&$\alpha$3\\	
		$^{274}$Nh&	11.668& 0.674 $\times$  $10^{-03}$&  0.568 $\times$  $10^{-04}$&  0.142 $\times$  $10^{-03}$&   0.323 $\times$  $10^{-03}$&0.209 $\times$  $10^{-04}$&-1.424&SF\\	
		$^{270}$Rg&	11.262& 0.152 $\times$  $10^{-02}$&  0.119 $\times$  $10^{-03}$&  0.305 $\times$  $10^{-03}$&   0.695 $\times$  $10^{-03}$&0.403 $\times$  $10^{-04}$&-2.535&SF\\	
		&&&&&&&&\\
		$^{287}119$&	13.795& 0.553 $\times$  $10^{-06}$&  0.229 $\times$  $10^{-06}$&  0.295 $\times$  $10^{-06}$&  0.210 $\times$  $10^{-06}$& 0.310 $\times$  $10^{-07}$&12.337&$\alpha$1\\	
		$^{283}$Ts&	13.092& 0.379 $\times$  $10^{-05}$&  0.118 $\times$  $10^{-05}$&  0.188 $\times$  $10^{-05}$&  0.138 $\times$  $10^{-05}$& 0.151 $\times$  $10^{-06}$& 7.224&$\alpha$2\\	
		$^{279}$Mc&	12.296& 0.470 $\times$  $10^{-04}$&  0.107 $\times$  $10^{-04}$&  0.219 $\times$  $10^{-04}$&  0.165 $\times$  $10^{-04}$& 0.124 $\times$  $10^{-05}$& 3.462&$\alpha$3\\	
		$^{275}$Nh&	11.629& 0.376 $\times$  $10^{-03}$&  0.681 $\times$  $10^{-04}$&  0.167 $\times$  $10^{-03}$&  0.128 $\times$  $10^{-03}$& 0.699 $\times$  $10^{-05}$& 0.974&$\alpha$4\\	
		$^{271}$Rg&	11.077& 0.188 $\times$  $10^{-02}$&  0.295 $\times$  $10^{-03}$&  0.797 $\times$  $10^{-03}$&  0.631 $\times$  $10^{-03}$& 0.268 $\times$  $10^{-04}$&-0.318&SF\\	
		&&&&&&&&\\
		$^{288}119$&	13.939& 0.537 $\times$  $10^{-06}$&    0.112 $\times$  $10^{-06}$&   0.126 $\times$  $10^{-06}$& 0.207 $\times$  $10^{-06}$& 0.548 $\times$  $10^{-07}$&14.401&$\alpha$1\\	
		$^{284}$Ts&	12.970& 0.145 $\times$  $10^{-04}$&    0.193 $\times$  $10^{-05}$&   0.314 $\times$  $10^{-05}$& 0.615 $\times$  $10^{-05}$& 0.865 $\times$  $10^{-06}$& 9.184&$\alpha$2\\	
		$^{280}$Mc&	12.042& 0.363 $\times$  $10^{-03}$&    0.328 $\times$  $10^{-04}$&   0.737 $\times$  $10^{-04}$& 0.171 $\times$  $10^{-03}$& 0.129 $\times$  $10^{-04}$& 5.310&$\alpha$3\\	
		$^{276}$Nh&	11.541& 0.131 $\times$  $10^{-02}$&    0.103 $\times$  $10^{-03}$&   0.252 $\times$  $10^{-03}$& 0.604 $\times$  $10^{-03}$& 0.370 $\times$  $10^{-04}$& 2.703&$\alpha$4\\	
		$^{272}$Rg&	10.971& 0.744 $\times$  $10^{-02}$&    0.502 $\times$  $10^{-03}$&   0.137 $\times$  $10^{-02}$& 0.351 $\times$  $10^{-02}$& 0.158 $\times$  $10^{-03}$& 1.284&$\alpha$5\\	
		$^{268}$Mt&	10.409& 4.580 $\times$  $10^{-02}$&    2.730 $\times$  $10^{-03}$&   8.190 $\times$  $10^{-03}$& 2.240 $\times$  $10^{-02}$& 7.280 $\times$  $10^{-04}$& 0.967&$\alpha$6\\
		$^{264}$Bh&	9.358&  8.010 $\times$  $10^{+00}$&    3.350 $\times$  $10^{-01}$&   1.410 $\times$  $10^{+00}$& 5.260 $\times$  $10^{+00}$& 6.000 $\times$  $10^{-02}$& 1.665&$\alpha$7/SF\\
		$^{260}$Db&	8.543&  5.630 $\times$  $10^{+02}$&    1.930 $\times$  $10^{+01}$&   9.860 $\times$  $10^{+01}$& 4.720 $\times$  $10^{+02}$& 2.290 $\times$  $10^{+00}$& 3.282&SF\\
		$^{256}$Lr&	7.690&  1.050 $\times$  $10^{+05}$&    3.020 $\times$  $10^{+03}$&   1.880 $\times$  $10^{+04}$& 1.230 $\times$  $10^{+05}$& 2.060 $\times$  $10^{+02}$& 5.718&SF\\
		&&&&&&&&\\
		$^{289}119$&	14.116& 0.145 $\times$  $10^{-06}$& 0.706 $\times$  $10^{-07}$&   0.714 $\times$  $10^{-07}$&   0.504 $\times$  $10^{-07}$& 0.982 $\times$  $10^{-08}$&15.650&$\alpha$1\\	
		$^{285}$Ts&	13.015& 0.537 $\times$  $10^{-05}$& 0.161 $\times$  $10^{-05}$&   0.246 $\times$  $10^{-05}$&   0.179 $\times$  $10^{-05}$& 0.205 $\times$  $10^{-06}$&10.379&$\alpha$2\\	
		$^{281}$Mc&	11.770& 0.665 $\times$  $10^{-03}$& 0.114 $\times$  $10^{-03}$&   0.283 $\times$  $10^{-03}$&   0.216 $\times$  $10^{-03}$& 0.119 $\times$  $10^{-04}$& 6.445&$\alpha$3\\	
		$^{277}$Nh&	11.462& 0.904 $\times$  $10^{-03}$& 0.149 $\times$  $10^{-03}$&   0.367 $\times$  $10^{-03}$&   0.284 $\times$  $10^{-03}$& 0.149 $\times$  $10^{-04}$& 3.770&$\alpha$4\\	
		$^{273}$Rg&	10.931& 0.423 $\times$  $10^{-02}$& 0.615 $\times$  $10^{-03}$&   0.165 $\times$  $10^{-02}$&	 0.131 $\times$  $10^{-02}$& 0.537 $\times$  $10^{-04}$& 2.278&$\alpha$5\\	
		$^{269}$Mt&	10.211& 6.840 $\times$  $10^{-02}$& 8.070 $\times$  $10^{-03}$&   2.580 $\times$  $10^{-02}$&   2.110 $\times$  $10^{-02}$& 5.670 $\times$  $10^{-04}$& 1.882&$\alpha$6\\
		$^{265}$Bh&	9.216&	 9.690 $\times$  $10^{+00}$& 8.220 $\times$  $10^{-01}$&   3.590 $\times$  $10^{+00}$&   3.060 $\times$  $10^{+00}$& 3.900 $\times$  $10^{-02}$& 2.494&$\alpha$7/SF\\
		$^{261}$Db&	8.316&	 1.510 $\times$  $10^{+03}$& 9.990 $\times$  $10^{+01}$&   5.580 $\times$  $10^{+02}$&   4.970 $\times$  $10^{+02}$& 2.950 $\times$  $10^{+00}$& 4.019& SF\\
		$^{257}$Lr&	7.597&  1.090 $\times$  $10^{+05}$& 6.530 $\times$  $10^{+03}$&   4.120 $\times$  $10^{+04}$&   3.800 $\times$  $10^{+04}$& 1.180 $\times$  $10^{+02}$& 6.356& SF\\
		&&&&&&&&\\
		$^{290}119$&	14.219& 0.208 $\times$  $10^{-06}$&   0.487 $\times$  $10^{-07}$&  0.451 $\times$  $10^{-07}$&	  0.693 $\times$  $10^{-07}$& 0.244 $\times$  $10^{-07}$&16.093&$\alpha$1\\	
		$^{286}$Ts&	13.114& 0.754 $\times$  $10^{-05}$&   0.108 $\times$  $10^{-05}$&  0.151 $\times$  $10^{-05}$&	  0.281 $\times$  $10^{-05}$& 0.494 $\times$  $10^{-06}$&10.817&$\alpha$2\\	
		$^{282}$Mc&	11.617& 0.327 $\times$  $10^{-02}$&   0.232 $\times$  $10^{-03}$&  0.606 $\times$  $10^{-03}$&	  0.165 $\times$  $10^{-02}$& 0.848 $\times$  $10^{-04}$& 6.872&$\alpha$3\\	
		$^{278}$Nh&	11.057 (11.60)$^*$& 0.182 $\times$  $10^{-01}$&   0.109 $\times$  $10^{-02}$&  0.321 $\times$  $10^{-02}$&	  0.926 $\times$  $10^{-02}$& 0.352 $\times$  $10^{-03}$& 4.183&$\alpha$4\\	
		$^{274}$Rg&	10.975 (11.15)$^*$& 0.727 $\times$  $10^{-02}$&   0.492 $\times$  $10^{-03}$&  0.124 $\times$  $10^{-02}$&	  0.315 $\times$  $10^{-02}$& 0.155 $\times$  $10^{-03}$& 2.671&$\alpha$5\\	
		$^{270}$Mt&	10.282 (10.03)$^*$& 9.780 $\times$  $10^{-02}$&   5.450 $\times$  $10^{-03}$&  1.610 $\times$  $10^{-02}$&	  4.650 $\times$  $10^{-02}$& 1.400 $\times$  $10^{-03}$& 2.250&$\alpha$6\\
		$^{266}$Bh&	8.901&	 2.020 $\times$  $10^{+02}$&   6.530 $\times$  $10^{+00}$&  3.280 $\times$  $10^{+01}$&	  1.540 $\times$  $10^{+02}$& 9.610 $\times$  $10^{-01}$& 2.833&$\alpha$7/SF\\
		$^{262}$Db&	8.132&	 1.470 $\times$  $10^{+04}$&   3.980 $\times$  $10^{+02}$&  2.380 $\times$  $10^{+03}$&	  1.440 $\times$  $10^{+04}$& 3.780 $\times$  $10^{+01}$& 4.325&SF\\
		$^{258}$Lr&	7.446&	 9.470 $\times$  $10^{+05}$&   2.350 $\times$  $10^{+04}$&  1.560 $\times$  $10^{+05}$&	  1.200 $\times$  $10^{+06}$& 1.360 $\times$  $10^{+03}$& 6.625&SF\\
		&&&&&&&&\\
		
		\hline\hline
		\label{tab5}
	\end{tabular}
\end{table*}

\begin{table*}
	\caption{Table 5 is continued.....}
	\renewcommand{\tabcolsep}{0.15cm}
	\renewcommand{\arraystretch}{1.0}
	\footnotesize\footnotesize
	\begin{tabular}{ccccccccc}
		\hline\hline
		Nuclei&$Q_\alpha^{RMF}$&\multicolumn{5}{c}{$T_{1/2}^\alpha$}&{$T_{1/2}^{SF}$}&Mode of \\
		\cline{3-7} 
		&&VSS&Brown&Royer&GLDM&Ni et. al.&Ren-Xu&decay\\
		\hline
		$^{291}119$&	14.378& 0.499 $\times$  $10^{-07}$&   0.278 $\times$  $10^{-07}$&  0.228 $\times$  $10^{-07}$&	 	0.160 $\times$  $10^{-07}$& 0.396 $\times$  $10^{-08}$&	15.739&$\alpha$1\\	
		$^{287}$Ts&	13.071& 0.417 $\times$  $10^{-05}$&   0.129 $\times$  $10^{-05}$&  0.175 $\times$  $10^{-05}$&	 	0.128 $\times$  $10^{-05}$& 0.164 $\times$  $10^{-06}$&	10.506&$\alpha$2\\	
		$^{283}$Mc&	11.570& 0.419 $\times$  $10^{-02}$&   0.291 $\times$  $10^{-03}$&  0.778 $\times$  $10^{-03}$&	 	0.215 $\times$  $10^{-02}$& 0.105 $\times$  $10^{-03}$&	 6.523&$\alpha$3\\	
		$^{279}$Nh&	10.712& 0.132 $\times$  $10^{+00}$&   0.649 $\times$  $10^{-02}$&  0.232 $\times$  $10^{-01}$&	 	0.773 $\times$  $10^{-01}$& 0.193 $\times$  $10^{-02}$&	 3.859&$\alpha$4\\	
		$^{275}$Rg&	10.962& 0.782 $\times$  $10^{-02}$&   0.525 $\times$  $10^{-03}$&  0.133 $\times$  $10^{-02}$&	  	 0.340 $\times$  $10^{-02}$&  0.165 $\times$  $10^{-03}$&	 2.374&$\alpha$5\\	
		$^{271}$Mt&	10.320& 3.540 $\times$  $10^{-02}$&   4.430 $\times$  $10^{-03}$&  1.230 $\times$  $10^{-02}$&  	 9.970 $\times$  $10^{-03}$& 3.220 $\times$  $10^{-04}$&	 2.079&$\alpha$6\\
		$^{267}$Bh&	8.729&  3.320 $\times$  $10^{+02}$&   2.120 $\times$  $10^{+01}$&  1.130 $\times$  $10^{+02}$&   	 9.810 $\times$  $10^{+01}$& 8.110 $\times$  $10^{-01}$&	 2.689&SF\\
		$^{263}$Db&	7.786&  1.270 $\times$  $10^{+05}$&   6.110 $\times$  $10^{+03}$&  4.370 $\times$  $10^{+04}$&   	 3.970 $\times$  $10^{+04}$& 1.330 $\times$  $10^{+02}$&	 4.206&SF\\
		$^{259}$Lr&	7.334&  1.220 $\times$  $10^{+06}$&   6.250 $\times$  $10^{+04}$&  4.260 $\times$  $10^{+05}$&   	 3.980 $\times$  $10^{+05}$& 9.370 $\times$  $10^{+02}$&	 6.528&SF\\
		$^{292}119$&14.450&0.828 $\times$  $10^{-07}$&   0.217 $\times$ $10^{-07}$& 0.165 $\times$ $10^{-07}$& 0.238 $\times$ $10^{-07}$& 0.111 $\times$ $10^{-07}$& 14.595&$\alpha$1\\ 	
		$^{288}$Ts&13.081& 0.875 $\times$  $10^{-05}$&   0.124 $\times$ $10^{-05}$& 0.161 $\times$ $10^{-05}$& 0.304 $\times$ $10^{-05}$& 0.561 $\times$ $10^{-06}$& 9.454&$\alpha$2\\	
		$^{284}$Mc&11.552& 0.462 $\times$  $10^{-02}$&   0.317 $\times$ $10^{-03}$& 0.788 $\times$ $10^{-03}$& 0.219 $\times$ $10^{-02}$& 0.114 $\times$ $10^{-03}$& 5.638&$\alpha$3\\	
		$^{280}$Nh&10.486& 0.510 $\times$  $10^{-00}$&   0.218 $\times$ $10^{-01}$& 0.823 $\times$ $10^{-01}$& 0.302 $\times$ $10^{-00}$& 0.614 $\times$ $10^{-02}$& 3.073&$\alpha$4\\	
		$^{276}$Rg&10.434& 0.166 $\times$  $10^{-00}$&   0.836 $\times$ $10^{-02}$& 0.259 $\times$ $10^{-01}$& 0.827 $\times$ $10^{-01}$& 0.227 $\times$ $10^{-02}$& 1.679&$ \alpha$5\\	
		$^{272}$Mt&10.560& 1.890 $\times$  $10^{-02}$&   1.220 $\times$ $10^{-03}$& 2.860 $\times$ $10^{-03}$& 7.360 $\times$ $10^{-03}$& 3.410 $\times$ $10^{-04}$& 1.374& $\alpha$6\\
		$^{268}$Bh& 8.804& 4.150 $\times$  $10^{+02}$&   1.260 $\times$ $10^{+01}$& 6.180 $\times$ $10^{+01}$& 3.070 $\times$ $10^{+02}$& 1.780 $\times$ $10^{+00}$& 2.067& SF\\
		$^{264}$Db& 7.442& 6.400 $\times$  $10^{+06}$&   1.110 $\times$ $10^{+05}$& 9.640 $\times$ $10^{+05}$& 9.010 $\times$ $10^{+06}$& 6.990 $\times$ $10^{+03}$& 3.665& SF\\
		$^{260}$Lr& 7.144& 1.670 $\times$  $10^{+07}$&   3.450 $\times$ $10^{+05}$& 2.550 $\times$ $10^{+06}$& 2.410 $\times$ $10^{+07}$& 1.600 $\times$ $10^{+04}$& 6.069& SF\\
		&&&&&&&&\\
		$^{293}119$&15.317&0.139 $\times$  $10^{-08}$&   0.120 $\times$ $10^{-08}$& 0.596 $\times$ $10^{-09}$& 0.409 $\times$ $10^{-09}$& 0.185 $\times$ $10^{-09}$& 12.670&$\alpha$1\\	
		$^{289}$Ts&12.984& 0.619 $\times$  $10^{-05}$&   0.183 $\times$ $10^{-05}$& 0.240 $\times$ $10^{-05}$& 0.174 $\times$ $10^{-05}$& 0.230 $\times$ $10^{-06}$& 7.668&$\alpha$2\\	
		$^{285}$Mc&11.603& 0.160 $\times$  $10^{-02}$&   0.248 $\times$ $10^{-03}$& 0.577 $\times$ $10^{-03}$& 0.439 $\times$ $10^{-03}$& 0.254 $\times$ $10^{-04}$& 3.990&$\alpha$3\\	
		$^{281}$Nh&10.317& 0.656 $\times$  $10^{+00}$&   0.556 $\times$ $10^{-01}$& 0.223 $\times$ $10^{+00}$& 0.178 $\times$ $10^{+00}$& 0.421 $\times$ $10^{-02}$& 1.563&$ \alpha$4\\	
		$^{277}$Rg&10.169& 0.384 $\times$  $10^{+00}$&   0.363 $\times$ $10^{-01}$& 0.126 $\times$ $10^{+00}$& 0.102 $\times$ $10^{+00}$& 0.256 $\times$ $10^{-02}$& 0.308&$ \alpha$5\\	
		$^{273}$Mt&10.508& 1.160 $\times$  $10^{-02}$&   1.610 $\times$ $10^{-03}$& 3.710 $\times$ $10^{-03}$& 2.990 $\times$ $10^{-03}$& 1.240 $\times$ $10^{-04}$& 0.140&$\alpha$6\\
		$^{269}$Bh& 8.814& 1.750 $\times$  $10^{-02}$&   1.180 $\times$ $10^{+01}$& 5.500 $\times$ $10^{+01}$& 4.740 $\times$ $10^{-01}$& 4.690 $\times$ $10^{-01}$& 0.971&$\alpha$7/SF\\
		$^{261}$Db& 6.833& 1.780 $\times$  $10^{+08}$&   6.580 $\times$ $10^{+06}$& 5.780 $\times$ $10^{+07}$& 5.530 $\times$ $10^{+07}$& 6.780 $\times$ $10^{+04}$& 5.253&SF\\
		&&&&&&&&\\
		$^{294}119$&15.131&0.607 $\times$  $10^{-08}$&   0.218 $\times$ $10^{-08}$& 0.113 $\times$ $10^{-08}$& 0.135 $\times$ $10^{-08}$& 0.118 $\times$ $10^{-08}$& 9.971&$\alpha$1\\	
		$^{290}$Ts&12.956& 0.155 $\times$  $10^{-04}$&   0.204 $\times$ $10^{-05}$& 0.261 $\times$ $10^{-05}$& 0.514 $\times$ $10^{-05}$& 0.914 $\times$ $10^{-06}$& 5.156&$\alpha$2\\	
		$^{286}$Mc&11.561& 0.440 $\times$  $10^{-02}$&   0.303 $\times$ $10^{-03}$& 0.692 $\times$ $10^{-03}$& 0.192 $\times$ $10^{-02}$& 0.110 $\times$ $10^{-03}$& 1.666&$\alpha$3\\	
		$^{282}$Nh&10.188& 0.325 $\times$  $10^{+01}$&   0.115 $\times$ $10^{+00}$& 0.481 $\times$ $10^{+00}$& 0.201 $\times$ $10^{+01}$& 0.300 $\times$ $10^{-01}$&-0.573&SF \\	
		$^{278}$Rg& 9.925& 0.398 $\times$  $10^{+01}$&   0.148 $\times$ $10^{+00}$& 0.570 $\times$ $10^{+00}$& 0.228 $\times$ $10^{+01}$& 0.345 $\times$ $10^{-01}$& 1.639& SF \\	
		&&&&&&&&\\
		$^{295}119$&16.177&0.695 $\times$ $10^{-10}$&    0.868 $\times$ $10^{-10}$& 0.277 $\times$ $10^{-10}$& 0.187 $\times$ $10^{-10}$& 0.142 $\times$ $10^{-10}$& 6.507& $\alpha$1\\
		$^{291}$Ts&11.763& 0.253 $\times$ $10^{-02}$&    0.372 $\times$ $10^{-03}$& 0.884 $\times$ $10^{-03}$& 0.663 $\times$ $10^{-03}$& 0.396 $\times$ $10^{-04}$& 1.925&  $\alpha$2\\	
		$^{287}$Mc&11.332 (10.74)$^*$& 0.695 $\times$ $10^{-02}$&    0.918 $\times$ $10^{-03}$& 0.229 $\times$ $10^{-02}$& 0.176 $\times$ $10^{-02}$& 0.892 $\times$ $10^{-04}$& -1.329&SF\\	
		$^{283}$Nh&10.097 (10.26)$^*$& 0.581 $\times$ $10^{+01}$&    0.194 $\times$ $10^{+00}$& 0.859 $\times$ $10^{+00}$& 0.375 $\times$ $10^{+01}$& 0.494 $\times$ $10^{-01}$& -3.443& SF\\	
		$^{279}$Rg& 9.666 (10.52)$^*$& 0.221 $\times$ $10^{+02}$&    0.697 $\times$ $10^{+00}$& 0.315 $\times$ $10^{+01}$& 0.143 $\times$ $10^{+02}$& 0.150 $\times$ $10^{+00}$& -4.273& SF \\	
		&&&&&&&&\\
		$^{296}119$&16.017&0.262 $\times$ $10^{-09}$&    0.139 $\times$ $10^{-09}$& 0.455 $\times$ $10^{-10}$& 0.435 $\times$ $10^{-10}$& 0.803 $\times$ $10^{-10}$& 2.285&$\alpha$1 \\
		$^{292}$Ts&11.596& 0.136 $\times$ $10^{-01}$&    0.822 $\times$ $10^{-03}$& 0.207 $\times$ $10^{-02}$& 0.662 $\times$ $10^{-02}$& 0.304 $\times$ $10^{-03}$& -2.016&    SF   \\
		$^{288}$Mc&11.262 (10.46)$^*$& 0.225 $\times$ $10^{-01}$&    0.130 $\times$ $10^{-02}$& 0.324 $\times$ $10^{-02}$& 0.101 $\times$ $10^{-01}$& 0.443 $\times$ $10^{-03}$& -4.987&    SF  \\	
		$^{284}$Nh& 9.920 (10.00)$^*$& 0.184 $\times$ $10^{+02}$&    0.547 $\times$ $10^{+00}$& 0.250 $\times$ $10^{+01}$& 0.119 $\times$ $10^{+02}$& 0.133 $\times$ $10^{+00}$& -6.701& SF \\	
		$^{280}$Rg& 9.454 (9.75)$^*$& 0.943 $\times$ $10^{+02}$&    0.260 $\times$ $10^{+01}$& 0.124 $\times$ $10^{+02}$& 0.622 $\times$ $10^{+02}$& 0.521 $\times$ $10^{+00}$& -7.235& SF \\
		\hline\hline
		\label{tab6}
	\end{tabular}
\end{table*}
\begin{table*}
	\caption{ CPPM study of $\alpha-$decay chains of fission survival 
		nuclides (i.e.$^{284-290}119$). 
		Experimental data for $Q_\alpha$~\cite{expq1,expq2,hamilton2012}, 
	if available, is given in parentheses with asterisk.
	}
	\renewcommand{\tabcolsep}{0.40cm}
	\renewcommand{\arraystretch}{0.9}
	\footnotesize\footnotesize
	\begin{tabular}{cccccc}
		\hline\hline
		Nuclei&$Q_\alpha^{RMF}$&\multicolumn{2}{c}{$T_{1/2}^\alpha$}&{$T_{1/2}^{SF}$}&Mode of \\
		\cline{3-4} 
		&&CPPM ( RMF Q$_{\alpha}$ )&CPPM (expt. Q$_{\alpha}$)&Santhosh et al&decay\\
		\hline
		$^{284}119$&	14.005& 1.17 $\times$  $10^{-07}$& & 8.47 $\times$ 10$^{6}$&$\alpha$\\
					
		$^{280}$Ts&	13.257& 1.05 $\times$  $10^{-06}$&  & 9.61 $\times $10$^{2} $&$\alpha$\\
		
		$^{276}$Mc&	12.353& 2.49 $\times$  $10^{-05}$&  & 3.96 $\times $10$^{-1} $& $\alpha$\\
			
		$^{272}$Nh&	12.032& 3.29 $\times$  $10^{-05}$&  & 3.99 $\times $10$^{-3} $& $\alpha$\\
		$^{268}$Rg&	11.714& 4.33 $\times$  $10^{-05}$&  & 3.31 $\times $10$^{-4} $& $\alpha$ \\
        &&&&&\\
        $^{285}119$&	13.891& 1.94 $\times$  $10^{-07}$& & 6.27 $\times$ 10$^{7}$&$\alpha$\\
        
        $^{281}$Ts&	13.396& 5.20 $\times$  $10^{-07}$&  & 2.30$\times $10$^{04} $&$\alpha$\\
        
        $^{277}$Mc&	12.286& 3.39 $\times$  $10^{-05}$&  & 5.12 $\times $10$^{00} $& $\alpha$\\
        
        $^{273}$Nh&	11.873& 7.44 $\times$  $10^{-05}$&  & 6.11 $\times $10$^{-02} $& $\alpha$\\
        $^{269}$Rg&	11.398& 2.43 $\times$  $10^{-04}$&  & 4.58 $\times $10$^{-03} $& $\alpha$ \\
         &&&&&\\
        
         $^{286}119$&	13.827& 2.45 $\times$  $10^{-07}$& & 7.40 $\times$ 10$^{+08}$&$\alpha$\\
        
        $^{282}$Ts&	13.296& 8.02 $\times$  $10^{-07}$&  & 2.15$\times $10$^{+05} $&$\alpha$\\
        
        $^{278}$Mc&	12.306& 2.93 $\times$  $10^{-05}$&  & 1.22 $\times $10$^{+02} $& $\alpha$\\
        
        $^{274}$Nh&	11.668& 2.22 $\times$  $10^{-04}$&  & 1.11 $\times $10$^{+00} $& $\alpha$\\
        $^{270}$Rg&	11.262& 5.08$\times$  $10^{-04}$&  & 9.90 $\times $10$^{-02} $& $\alpha$ \\
        &&&&&\\
        
         $^{287}119$&	13.795& 2.82 $\times$  $10^{-07}$& & 5.81 $\times$ 10$^{+09}$&$\alpha$\\
        
        $^{283}$Ts&	13.092& 2.07$\times$  $10^{-06}$&  & 1.09$\times $10$^{+06} $&$\alpha$\\
        
        $^{279}$Mc&	12.296& 2.99 $\times$  $10^{-05}$&  & 1.89$\times $10$^{+03} $& $\alpha$\\
        
        $^{275}$Nh&	11.629& 2.65 $\times$  $10^{-04}$&  & 8.04 $\times $10$^{+00} $& $\alpha$\\
        $^{271}$Rg&	11.077& 1.44$\times$  $10^{-03}$&  & 7.66 $\times $10$^{-01} $& $\alpha$ \\
        &&&&&\\

         $^{288}119$&	13.939& 1.36 $\times$  $10^{-07}$& & 2.02 $\times$ 10$^{+10}$&$\alpha$\\
        
        $^{284}$Ts&	12.97& 3.62$\times$  $10^{-06}$&  & 8.06$\times $10$^{+06} $&$\alpha$\\
        
        $^{280}$Mc&	12.042& 1.11 $\times$  $10^{-04}$&  & 1.06$\times $10$^{+04} $& $\alpha$\\
        
        $^{276}$Nh&	11.541& 4.18 $\times$  $10^{-04}$&  & 5.60 $\times $10$^{+01} $& $\alpha$\\
        $^{272}$Rg&	10.971& 2.61$\times$  $10^{-03}$&  & 7.50 $\times $10$^{+00} $& $\alpha$ \\
        
        $^{268}$Mt&	10.409& 1.80$\times$  $10^{-02}$&  & 2.80$\times $10$^{+00} $&$\alpha$\\
        
        $^{264}$Bh&	9.358& 4.59852 &  & 2.80$\times $10$^{+00} $& SF\\
        
        $^{260}$Db&	8.543& 443.74 &  & 2.30 $\times $10$^{+01} $& SF\\
        $^{256}$Lr&	7.69& 123998.0&  & 1.99 $\times $10$^{+03} $& SF \\
         &&&&&\\
         
          $^{289}119$&	14.116& 5.83 $\times$  $10^{-08}$& & 4.81 $\times$ 10$^{+10}$&$\alpha$\\
         
         $^{285}$Ts&	13.015&2.78 $\times$  $10^{-06}$&  & 2.27$\times $10$^{+07} $&$\alpha$\\
         
         $^{281}$Mc&	11.77&4.78 $\times$  $10^{-04}$&  & 2.76$\times $10$^{+04} $& $\alpha$\\
         
         $^{277}$Nh&	11.462& 6.33 $\times$  $10^{-04}$&  & 9.11 $\times $10$^{+01} $& $\alpha$\\
         $^{273}$Rg&	10.931& 3.20$\times$  $10^{-03}$&  & 2.90 $\times $10$^{+01} $& $\alpha$ \\
         
         $^{269}$Mt&	10.211& 6.23$\times$  $10^{-02}$&  & 1.04$\times $10$^{+01} $&$\alpha$\\
         
         $^{265}$Bh&	9.216& 12.6775 &  & 7.14$\times $10$^{+00} $& SF\\
         
         $^{261}$Db&	8.316& 2874.07&  & 3.47 $\times $10$^{+01} $& SF\\
         $^{257}$Lr&	7.597& 289343.0&  & 2.89 $\times $10$^{+03} $& SF \\
          &&&&&\\
         $^{290}119$&	14.219& 3.60 $\times$  $10^{-08}$& & 1.88 $\times$ 10$^{11}$&$\alpha$\\
         
         $^{286}$Ts&	13.114&1.64$\times$  $10^{-06}$&  & 8.98$\times $10$^{+07} $&$\alpha$\\
         
         $^{282}$Mc&	11.617&1.09 $\times$  $10^{-03}$&  & 8.65$\times $10$^{+04} $& $\alpha$\\
         
         $^{278}$Nh&	11.057 (11.60)& 6.61 $\times$  $10^{-03}$& 2.77 $\times$  $10^{-04}$ & 1.33 $\times $10$^{+02} $& $\alpha$\\
         $^{274}$Rg&	10.975(11.15)& 2.36$\times$  $10^{-03}$& 8.33$\times$  $10^{-04}$ & 1.01 $\times $10$^{+02} $& $\alpha$ \\
         
         $^{270}$Mt&	10.282(10.03)& 3.77$\times$  $10^{-02}$& 0.199806 & 4.68$\times $10$^{+01} $&$\alpha$\\
         
         $^{266}$Bh&	8.901& 137.725 &  & 3.12$\times $10$^{+01} $& SF\\
         
         $^{262}$Db&	8.132& 13740.4&  & 8.89 $\times $10$^{+01} $& SF\\
         $^{258}$Lr&	7.446& 1.22 $\times$ $10^{+06} $ &  & 4.48 $\times $10$^{+03} $& SF \\
         
         &&&&&\\

		\hline\hline
		\label{tab}
	\end{tabular}
\end{table*}

\begin{table*}
	\caption{CPPM study of $\alpha-$decay chains of fission survival 
		nuclides (i.e.$^{291-296}119$). Experimental data for $Q_\alpha$~\cite{expq1,expq2,hamilton2012}, 
		if available, is given in parentheses with asterisk.}
	\renewcommand{\tabcolsep}{0.35cm}
	\renewcommand{\arraystretch}{0.9}
	\footnotesize\footnotesize
	\begin{tabular}{cccccc}
		\hline\hline
		Nuclei&$Q_\alpha^{RMF}$&\multicolumn{2}{c}{$T_{1/2}^\alpha$}&{$T_{1/2}^{SF}$}&Mode of \\
		\cline{3-4} 
		&&CPPM ( RMF Q$_{\alpha}$ )&CPPM (expt. Q$_{\alpha}$)&Santhosh et al&decay\\
		\hline
	
	
		
		
		
		
		
	

		 $^{291}119$&	14.378& 1.72 $\times$  $10^{-08}$& & 2.01 $\times$ 10$^{+11}$&$\alpha$\\
		
		$^{287}$Ts&	13.071&1.95 $\times$  $10^{-06}$&  & 1.17$\times $10$^{+08} $&$\alpha$\\
			
			$^{283}$Mc&	11.57&1.38 $\times$  $10^{-03}$&  & 1.33$\times $10$^{+05} $& $\alpha$\\
			
			$^{279}$Nh&	10.712& 0.053809 &  & 2.14 $\times $10$^{+02} $& $\alpha$\\
			$^{275}$Rg&	10.962& 2.45$\times$  $10^{-03}$&  & 5.86 $\times $10$^{+01} $& $\alpha$ \\
			
			$^{271}$Mt&	10.32& 2.83$\times$  $10^{-02}$&  & 9.11$\times $10$^{+01} $&$\alpha$\\
			
			$^{267}$Bh&	8.729& 525.71 &  & 5.45$\times $10$^{+01} $& SF\\
			
			$^{263}$Db&	7.786& 313269.0&  & 1.00 $\times $10$^{+02} $& SF\\
			$^{259}$Lr&	7.334& 3.59$\times$10$^{+06}$ &  & 2.47 $\times$10$^{+03}$ & SF \\
				&&&&&\\
						
			$^{292}$119&	14.45& 1.22 $\times$  $10^{-08}$& & 4.05 $\times$ 10$^{+11}$&$\alpha$\\
			
			$^{288}$Ts&	13.081&1.79 $\times$  $10^{-06}$&  & 2.39$\times $10$^{+08} $&$\alpha$\\
			
			$^{284}$Mc&	11.552&1.47 $\times$  $10^{-03}$&  & 3.07$\times $10$^{+05} $& $\alpha$\\
			
			$^{280}$Nh&	10.486& 0.221739&  & 2.65 $\times $10$^{+02} $& $\alpha$\\
			$^{276}$Rg&	10.434& 0.063858&  & 3.98 $\times $10$^{+01} $& $\alpha$ \\
			
			$^{272}$Mt&	10.56& 5.93$\times$  $10^{-03}$&  & 1.42$\times $10$^{+02} $&$\alpha$\\
			
			$^{268}$Bh&	8.804& 276.127 &  & 1.24$\times $10$^{+02} $& SF\\
			
			$^{264}$Db&	7.442& 8.68$\times$  $10^{+06}$ &  & 1.90 $\times $10$^{+02} $& SF\\
			$^{260}$Lr&	7.144& 2.46$\times $10$^{+07}$ &  & 2.85 $\times $10$^{+03} $& SF \\
				&&&&&\\
			$^{293}$119&	15.317& 3.25 $\times$  $10^{-10}$& & 2.37 $\times$ 10$^{+11}$&$\alpha$\\
			
			$^{289}$Ts&	12.984&2.77 $\times$  $10^{-06}$&  & 1.38$\times $10$^{+08} $&$\alpha$\\
			
			$^{285}$Mc&	11.603&1.05 $\times$  $10^{-03}$&  & 2.24$\times $10$^{+05} $& $\alpha$\\
			
			$^{281}$Nh&	10.317& 0.65282&  & 2.20 $\times $10$^{+02} $& $\alpha$\\
			$^{277}$Rg&	10.169& 0.353285&  & 1.05 $\times $10$^{+01} $& $\alpha$ \\
			
			$^{273}$Mt&	10.508& 7.91$\times$  $10^{-03}$&  & 4.19$\times $10$^{+01} $&$\alpha$\\
			
			$^{269}$Bh&	8.814& 245.475 &  & 8.66$\times $10$^{+01} $& SF\\
			
			$^{265}$Db&	6.833& 5.86$\times$  $10^{+09}$ &  & 1.34 $\times $10$^{+02} $& SF\\
			&&&&&\\

			$^{294}$119&	15.131& 6.68 $\times$  $10^{-10}$& & 2.27 $\times$ 10$^{+11}$&$\alpha$\\
			
			$^{290}$Ts&	12.956&3.07 $\times$  $10^{-06}$&  & 1.55$\times $10$^{+08} $&$\alpha$\\
			
			$^{286}$Mc&	11.561&1.29 $\times$  $10^{-03}$&  & 2.63$\times $10$^{+05} $& $\alpha$\\
			
			$^{282}$Nh&	10.188& 1.5027&  & 2.93 $\times $10$^{+02} $& $\alpha$\\
			$^{278}$Rg&	9.925& 1.80751&  & 3.98 $\times $10$^{+00} $& $\alpha$ \\
			
			&&&&&\\
			$^{295}$119&	16.177& 1.17 $\times$  $10^{-11}$& & 4.59 $\times$ 10$^{+10}$&$\alpha$\\
			
			$^{291}$Ts&	11.763&1.69 $\times$  $10^{-03}$&  & 5.10$\times $10$^{+07} $&$\alpha$\\
			
			$^{287}$Mc&	11.332 (10.74)&4.75 $\times$  $10^{-03}$& 0.183249 & 9.51$\times $10$^{+04} $& $\alpha$\\
			
			$^{283}$Nh&	10.097(10.26)& 2.70304& 0.887961 & 1.37 $\times $10$^{+02} $& $\alpha$\\
			$^{279}$Rg&	9.666 (10.52)& 10.9634& 0.032815 & 7.76 $\times $10$^{-01} $& SF \\
			&&&&&\\
			
			$^{296}$119&	16.017& 2.02 $\times$  $10^{-11}$& & 1.71 $\times$ 10$^{+10}$&$\alpha$\\
			
			$^{292}$Ts&	11.596&4.27 $\times$  $10^{-03}$&  & 2.82$\times $10$^{+07} $&$\alpha$\\
			
			$^{288}$Mc&	11.262 (10.46)&6.94$\times$  $10^{-03}$& 1.10276 & 6.34$\times $10$^{+04} $& $\alpha$\\
			
			$^{284}$Nh&	9.920(10.00)& 8.99101& 5.11525& 1.06 $\times $10$^{+02} $& $\alpha$\\
			$^{280}$Rg&	9.454 (9.75)& 50.3186& 5.7732 & 5.10 $\times $10$^{-01} $& SF \\
					
			\hline\hline
			\label{tab}
		\end{tabular}
	\end{table*}

\subsection{Decay-energies and half-lives}
Superheavy nuclei are identified by $\alpha-$decay followed by 
spontaneous fission as we have mentioned earlier. 
Decay energy $Q_{\alpha}$ is the basic parameter to understand 
the $\alpha-$decay and used to calculate the half-lives. 
It is observed in alpha-emission and new nucleus is identified. 
The knowledge of $Q_{\alpha}$ of a nucleus gives a valuable information 
about its stability. 
Decay energy is estimated by knowing the binding energies of the parent 
and daughter nuclei and binding energy of $^{4}$He nucleus. 
Here, the binding energies are calculated using the most reliable 
framework of relativistic mean-field model. 
$Q_{\alpha}$ is used as a basic input for calculating the $\alpha-$decay half-life.
The quantity $Q_{\alpha}$ is estimated using the relation
\begin{equation}
Q_{\alpha}(N,Z) = BE(N-2,Z-2)+ BE(2,2)- BE(N,Z).
\end{equation}
Here, $BE(N,Z)$, $BE(N-2,Z-2)$, and $BE(2,2)$ are the binding energies of
the parent, daughter and $^{4}$He (BE = 28.296 MeV~\cite{audi2003}) 
with neutron number N and proton number Z.
The values of $Q_{\alpha}$ for ground-state to ground (i.e. prolate) is 
estimated from RMF binding energy and are given in the Tables~$3-6$. 
In order to predict the dominant mode of decay of considered chain, we make 
the calculations for $\alpha-$decay, $\beta-$decay and spontaneous fission 
using various empirical formulas and comparison of their life-times 
shall provide the required results about the mode of decay.
The alpha decay half-lives are estimated using various empirical formulas 
given in literature such as Viola-Seaborg (VSS)~\cite{VJS66}, Brown~\cite{B92}, 
Royer~\cite{R0}, generalized liquid drop model (GLDM)~\cite{DR07}, Ni et al.~\cite{NDK08}. 
Spontaneous fission half-lives are computed using the 
semi-empirical formula of Ren and Xu~\cite{RX05}.
Here, Fiset and Nix~\cite{FN72} empirical formula is used to calculate the 
$\beta-$decay half-lives.

\subsubsection{Alpha decay}  
With the even-even values available at hand, 
the $\alpha-$decay half-life of the isotopic chain under study is estimated
by Viola-Seaborg semi-empirical relation
\begin{equation}
\log_{10}T^{\alpha}_{1/2} = \frac{aZ-b}{\sqrt{Q_{\alpha}}} -(cZ+d )+h_{\log}.
\end{equation}
The values of the parameters a, b, c and d are taken from the recent 
modified parameterizations of Sobiczewski et al~\cite{SPC89}, which are
$a = 1.66175$, $b=8.5166$, $c=0.20$, $d=33.9069$. 
The $h_{\log}$ is the hindrance factor which takes the care of odd numbers 
of proton and neutron as given by Viola and Seaborg
\begin{equation}
h_{\log}= \left\{
\begin{array}{rl}
0.000 & even-even;\\
0.772 & odd-even;\\
1.066 &  even- odd; \\
1.114 & odd-odd.
\end{array} \right.
\end{equation}
There are also several other phenomenological formulas available 
in the literature by which the $\alpha-$decay half-lives is calculated.
The semi-empirical formula proposed by Brown~\cite{B92} for determining
the half-life of superheavy nuclei is given by
\begin{equation}
\log_{10}T^{\alpha}_{1/2} = 9.54 (Z-2)^{0.6}/\sqrt{Q_{\alpha}} - 51.37,
\end{equation}
where Z, the atomic number of parent nucleus and $Q_{\alpha}$ decay 
energy are only the input for this formula.
Moreover, another theoretical predictions 
for half-life for heavy and superheavy nuclei by employing a 
fitting procedure to a set of 373 alpha emitters was developed by 
Royer~\cite{R0} with an RMS deviation of 0.42, given as
\begin{equation}
\log_{10}T^{\alpha}_{1/2} = -26.06 -1.114A^{1/6}\sqrt{Z} + \frac{1.5837Z}{\sqrt{Q_{\alpha}}},
\end{equation}
where A and Z represent the mass number and charge number 
of the parent nuclei and $Q_{\alpha}$ represents the energy released 
during the reaction. Assuming a similar dependence on A, Z and $Q_{\alpha}$, 
the above equation was reformulated for a subset of 131 even-even nuclei 
and a relation was obtained with a RMS deviation of only 0.285, given, as
\begin{equation}
\log_{10}T_{1/2}^{\alpha} = -25.31 -1.1629A^{1/6} \sqrt{Z} + \frac{1.5864Z}{\sqrt{Q_{\alpha}}}.
\end{equation}
For a subset of 106 even-odd nuclei, the relation given by was further modified 
with an RMS deviation of 0.39, and is given as,
\begin{equation}
\log_{10}T_{1/2}^{\alpha} = -26.65 -1.0859A^{1/6} \sqrt{Z} + \frac{1.5848Z}{\sqrt{Q_{\alpha}}}.
\end{equation}
A similar reformulation was performed for the equation for a 
subset of 86 odd-even nuclei and 50 odd-odd nuclei.
Another formula for $\alpha-$decay half-lives based on 
generalized liquid drop model proposed by 
Dasgupta-Schubert and Reyes~\cite{DR07} is obtained by fitting the 
experimental half-lives for 373 alpha emitters, given as
\begin{equation}
\log_{10}T^{\alpha}_{1/2} = a + bA^{1/6}Z^{1/2} + cZ/Q_{\alpha}^{1/2}.  
\end{equation} 
The parameters a, b and c are given by
\begin{equation}
a,b,c= \left\{
\begin{array}{rl}
-25.31,-1.1629,1.5864 & even-even;\\
-26.65,-1.0859,1.5848 & even-odd;\\
-25.68,-1.1423,1.5920 & odd-even;\\ 
-29.48,-1.1130,1.6971 & odd-odd\;.
\end{array} \right.
\end{equation}
Recently, in Ref.~\cite{NDK08} Ni et. al. proposed a unified formula for 
determining the half-lives in alpha decay and cluster radioactivity.
The formula for alpha decay is written as
\begin{equation}
\log_{10}T^{\alpha}_{1/2} = 2a\sqrt{\mu} (Z-2) Q_{\alpha}^{-1/2} + b \sqrt{\mu}[2(Z-2)]^{-1/2} + c
\end{equation}
where, a, b, c are the constants and $\mu$ is define as 4(A-4)/A.

\begin{figure}
\centering
	\resizebox{0.70\textwidth}{!}{%
		\includegraphics{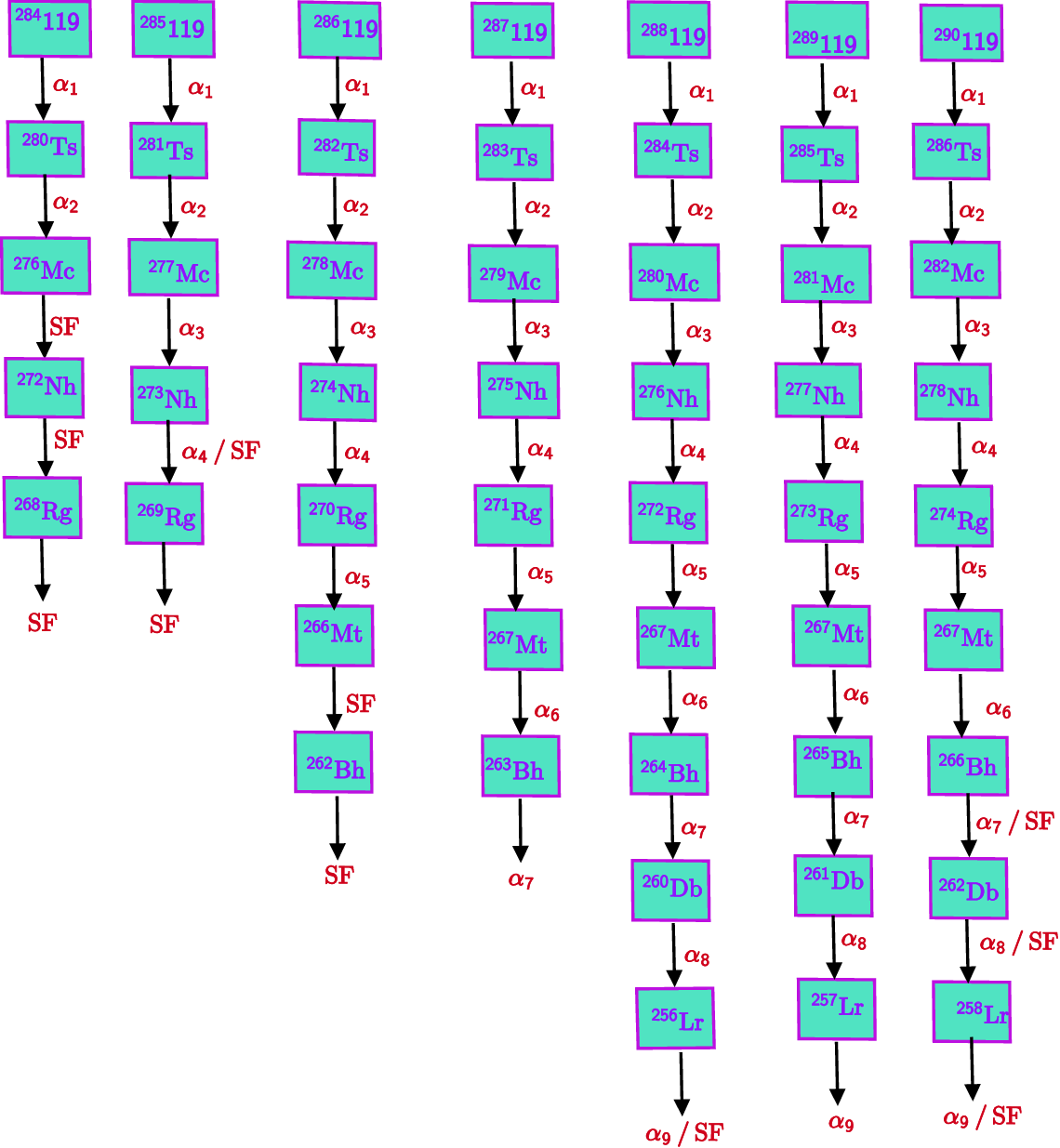}
	}
	\caption{(color online) Decay chain of $^{284-290}$119.}
	\label{Alphachains1}
\end{figure}

\begin{figure}
	\centering
	\resizebox{0.70\textwidth}{!}{%
		\includegraphics{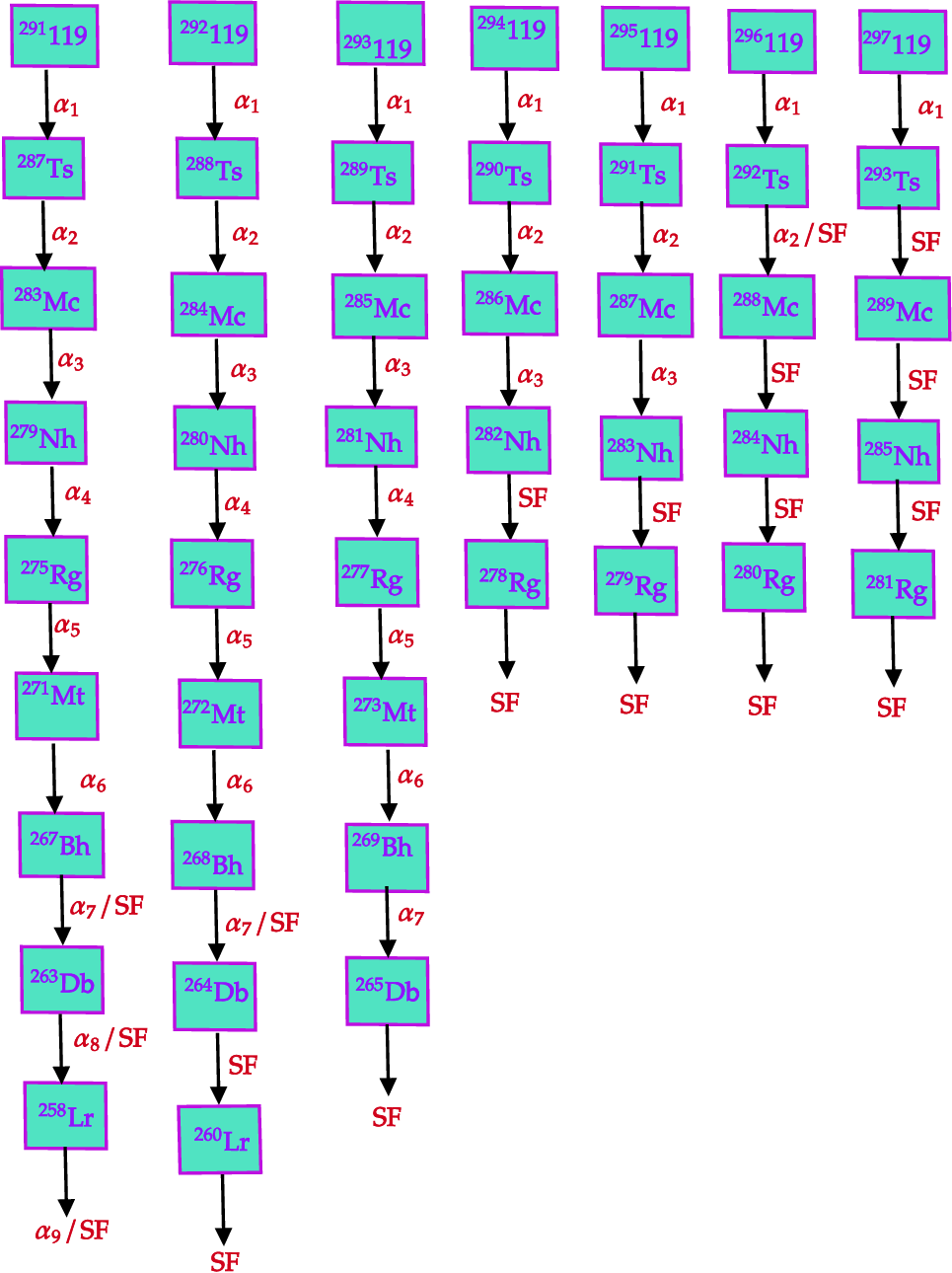}
	}
	\caption{(color online) Decay chain of $^{291-297}$119.}
	\label{Alphachains2}
\end{figure}

\subsubsection{Beta decay}
Beta decay, a three body decay mode,  is another very important mode of decay 
for the nuclei lie far from the stability line. 
The description of $\beta-$decay is explained by famous Fermi theory 
which describes the beta transition rates according to log(ft) values. 
It proceeds through weak interaction and this process is slow 
as well as less favoured compared to SF and alpha decay.
Recently, it is predicted that there may also be a possibility of $\beta-$decay in 
some of the superheavy nuclei where it may play a significant role~\cite{karpov2012}. 
In this regard, even in the presence of dominant mode of alpha and SF 
in SHN, we make the search for possibility of $\beta-$decay in order to 
completeness of decay modes of superheavy nuclei.   
To look out the possibility of $\beta-$decay in considered isotopic chain, we 
employed the empirical formula of Fiset and Nix~\cite{FN72} for estimating the half-lives 
of the isotopic chain under study which is given by
\begin{equation}
T_{1/2}^{\beta} = 540 \times 10^{5.0} \frac{m_{e}^{5}}{\rho_{d.o.s.}(W_{\beta}^{6}-m_{e}^{6})}.
\end{equation}
In an analogy of $\alpha-$decay, we evaluate the $Q_\beta$
value using the relation 
\begin{math}
Q_{\beta} = BE(Z+1,A)-BE(Z,A)
\end{math}
and further we calculate the $W_{\beta}$ by a relation
\begin{math} 
W_{\beta} = Q_{\beta} + m_{e}, 
\end{math}
where, $m_e$ is the rest mass of electron. 
Here, $\rho_{d.o.s.}$ is the average density of states in the
daughter nucleus ($e^{-A/290}\times$ number of states within 
1 MeV of ground state).

\subsubsection{Spontaneous Fission}
Superheavy nuclei are identified by alpha decay and the 
chain ends by spontaneous fission which helps in identifying the long
lived superheavy elements. 
Several empirical formulas for determining the spontaneous fission 
half-lives are available in literature proposed by various authors from 
time to time.

\paragraph{\textbf{Ren and Xu}}
In our calculations, we employed the phenomenological formula proposed by 
Ren and Xu ~\cite{RX05} expressed as
\begin{small}
	\begin{eqnarray}
	\log_{10}T^{SF}_{1/2}& = & 21.08 + C_{1} \frac{(Z-90-\nu)}{A}+C_{2}\frac{(Z-90-\nu)^{2}}{A}\nonumber\\
	&+& C_{3}\frac{(Z-90-\nu)^{3}}{A} \nonumber\\
	&+& C_{4} \frac{(Z-90-\nu)(N-Z-52)^{2}}{A} ,
	\end{eqnarray}
\end{small}
where Z, N, A represent the proton, neutron and mass number of parent nuclei.
C$_1$, C$_2$, C$_3$, C$_4$ are the empirical constants and $\nu$ is the seniority term 
which takes care of blocking effect of unpaired nucleons on the transfer 
of many nucleon pairs during the fission process.
\paragraph{\textbf{Santhosh et al}} \mbox{} \\
Spontaneous fission (SF) was first described within the geometrical framework of the charged liquid drop model by Bohr and Wheeler~\cite{BW39}. The first semi-empirical formula for finding the half-lives of SF was proposed by Swiatecki~\cite{S55} and showed that this formula can reproduce experimental values reasonably. The quantum tunnelling effect is considered as the underlying mechanism of SF and the probability of tunnelling depends exponentially on the square root of the barrier height and inversely proportional to the fissionability parameter $\frac{Z^2}{A}$. A semi-empirical formula for SF half-lives~\cite{SBS10} was developed by including fissionability parameter  and isospin effect $\frac{N-Z}{N+Z}$. Later this formula was modified~\cite{SN16} by incorporating shell correction term, as  shell structure plays an important role in determining SF half-lives and is given as 
     
\begin{small}
	\begin{equation}
	\log_{10}T^{SF}_{1/2}= a \frac{Z^2}{A}+b \bigg( \frac{Z^2}{A} \bigg)^2 + c \bigg(\frac{N-Z}{N+Z} \bigg)+ d \bigg(\frac{N-Z}{N+Z} \bigg)^2 + e E_{shell} + f
	\end{equation}
\end{small}
where a $= -43.25203$, b $= 0.49192$, c $ = 3674.3927$, d $= -9360.6$, e $= 0.8930$ and  f $= 578.56058$. E$_{shell}$ is the shell correction energy taken from Ref. ~\cite{MSIS16}.\\
We have studied the modes of decay of superheavy nuclei $^{284-296}$119 and their daughters in the alpha decay chains by comparing SF half-lives with corresponding alpha half-lives. The SF half-lives are computed using the formula of Santhosh et al~\cite{SN16} and alpha half-lives computed using CPPM~\cite{SJ00,SBSJ08}. For computing alpha half-lives Q values obtained using RMF formalism and in few cases available experimental Q values are used. The nuclei with alpha half-life less than SF half-life will undergo alpha decay and vice versa. The entire results of our calculations are given in Table 1. From the table it can be seen that our alpha half-life values computed using Q$_\alpha$(RMF) agree with half-life values computed using experimental Q$_\alpha$ values within 1 order difference. Also it can be seen that the isotopes $^{288-293}$119 exhibit consistent 6  chains followed by SF, isotopes $^{295,296}$119 exhibit consistent 4  chains followed by SF and the rest of the nuclei show continuous alpha chains. The isotopes $^{288-293}$119, $^{295,296}$119 will be of great interest to experimentalists for future studies.

\begin{figure}
\centering
	\resizebox{0.80\textwidth}{!}{%
		\includegraphics{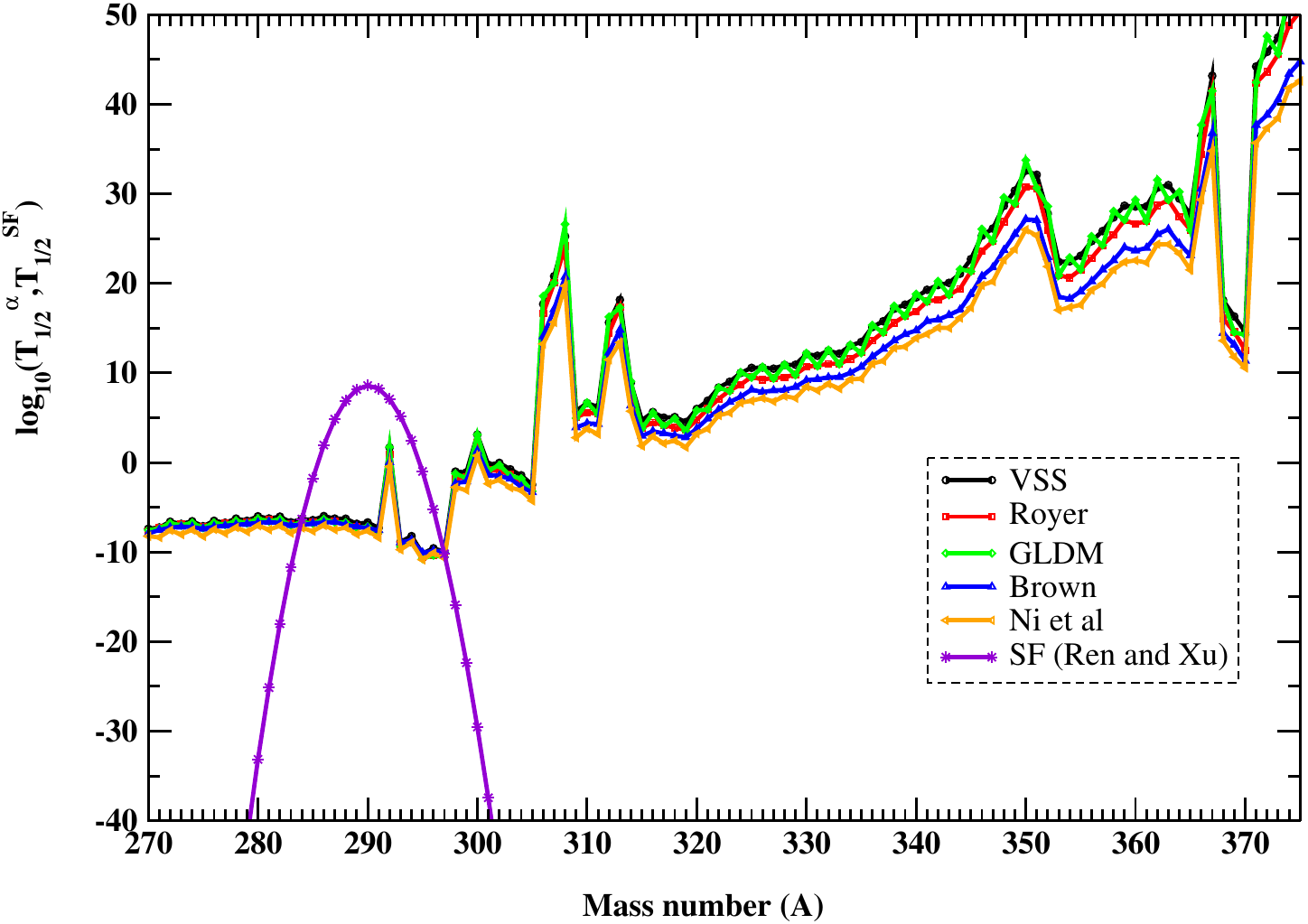}
	}
	\caption{(color online) Alpha decay and spontaneous fission half-lives of $Z=119$ 
		isotopic chain as a function of mass number.}
	\label{alife}
\end{figure}

Present analysis shows that some of the isotopes of $Z=119$ superheavy nuclei 
survived the fission and thus make the decay via $\alpha-$emission. 
The calculated $\alpha-$decay half-lives using VSS, Brown, Royer, GLDM and 
Ni et al. are framed in Tables~3, 4 and we noticed a good 
agreement among them as well as with FRDM data.
Fiset and Nix formula is employed to calculate the $\beta-$decay half-life for 
examining the possibility of mode of $\beta-$decay and the results are also 
presented in Tables~3 and 4.
It is noted that $\beta-$decay half-lives are found to be large than 
$\alpha-$decay as well as spontaneous fission half-lives and 
hence there is no possibility of mode of $\beta-$decay is observed 
for current isotopic chain.
Spontaneous fission half-lives is calculated using Ren and Xu formula 
and the estimated values are framed in one of the columns of Tables~$3-6$.
The calculated half-lives for $\alpha-$decay and SF are plotted against the 
mass number displayed in Fig. 5. 

Our calculations predict that the nuclides $^{284-296}$119 survive the 
fission and may be observed in the laboratory through alpha decay and the 
nuclei beyond A $>$ 296 do not survive fission and hence 
completely undergo spontaneous fission. 
Further, we aimed at predicting the possibility of $\alpha-$decay chain 
of fission survival nuclides i.e. $^{284-296}119$ of the considered isotopic 
chain given in Tables~5 and 6. 
Our study confirmed the possibility of one $\alpha$ chain 
from $^{284,296}119$, two consistent $\alpha$ chains from $^{285,295}119$, 
three consistent $\alpha$ chains from $^{286,294}119$, four consistent 
$\alpha$ chains from $^{287}119$, six consistent alpha chains from 
$^{288-293}119$ and these findings are reported in the Table 5 and 6. 
Unfortunately, there is no experimental information for $Z=119$ nuclides. 
But the experimental data of $Q_\alpha$ for a few decay elements of $Z=119$ 
is available~\cite{expq1,expq2,hamilton2012} and we mentioned in Tables 5 and 6. 
The calculated values of $Q_\alpha$ are compared with available experimental 
data~\cite{expq1,expq2,hamilton2012} and we found a close agreement between them. 
Moreover, the $\alpha-$decay chain of $^{295}119$ contains $^{291}$Ts,
$^{287}$Mc, $^{283}$Nh and $^{279}$Rg elements whose $\alpha-$decay 
chain is treated in Refs.~\cite{adamian2018,adamian2012} and a close 
agreement of our calculated $Q_\alpha$ with the values predicted 
in these Refs.~\cite{adamian2018,adamian2012} is noticed.
However, we did not mention the values of $Q_\alpha$ predicted in 
Refs.~\cite{adamian2018,adamian2012} into the manuscript. 
The inference drawn from this investigations is that the nuclides 
$^{284-296}119$ have the $\alpha-$decay chain with the life-time of the 
order of micro- or nano-second and thus these nuclides might be 
observed in the laboratory through alpha decay. 
We firmly believed that the alpha decay life-time of the 
isotopes $^{284-296}119$ presented in the manuscript may serve as a 
crucial theoretical input for designing the experimental setup and 
might provide a ray of hope in order to produce the yet-to-be 
synthesized isotopes of $Z=119$ in the laboratory in very near future.  

\section{Summary and Conclusion}
\label{summary}
In summary, we have calculated the structural properties
of $Z=119$ superheavy nuclei within a mass range $284 \leq A \leq 375$ 
using axially deformed relativistic mean field model. 
The calculations are performed for three different shape configurations 
prolate, oblate and spherical configurations in which prolate 
is suggested to be possible ground state for most of the nuclei.
Binding energy produced by RMF are in good agreement with FRDM data.
Two dimensional contour plot of density distribution has been made 
for predicting neutron shell closure nuclides $^{291}$119 and $^{303}$119 
to reveal the special features of the nuclei such as bubble or cluster structures.
Further, the predictions of possible modes of decay such as $\alpha-$decay, 
$\beta-$decay and spontaneous fission of the isotopic chain of $Z=119$ in 
the mass range 284 $\le$ A $\le$ 375 have been made.
The calculations performed for $\alpha$ decay half-lives
using the semi-empirical formulae Viola-Seaborg, Brown, Royer, GLDM 
and Ni et. al. are in good agreement with among each other 
as well as with macro-microscopic FRDM data wherever available. 
In addition, a thorough study on $\beta-$decay and SF half-lives 
have also been made to identify the mode of the decay of these isotopes. 
We conclude that the $\alpha-$decay and spontaneous fission are the 
principal modes of decay in considered chain of nuclides and there is no 
possibility of $\beta-$decay for the considered chain of nuclides under study.
The calculated values of $Q_\alpha$ are compared with experimental 
data~\cite{expq1,expq2,hamilton2012}, wherever available and found a 
close agreement between them. 
Moreover, our calculated $Q_\alpha$ are quite agreeable with the values 
predicted in Refs.~\cite{adamian2018,adamian2012}.
From our analysis we inferred  that the isotopes with mass number 284 $\le$ A $\le$ 296 
will survive fission and can be observed in the laboratory through alpha 
decay while beyond the mass number $A>296$ do not survive fission and 
hence completely undergoes spontaneous fission. 
We also analyzed the $\alpha-$ decay chain of fission survival nuclides i.e. 
$^{284-296}$119 for the considered isotopic chain and 
predicted one $\alpha$ decay chain for $^{284,296}$119 followed by SF, two consistent $\alpha$ decay chains for $^{285,295}$119, three consistent $\alpha$ decay chains  for $^{286,294}$119, four consistent $\alpha$ decay chains for $^{287}$119 and
six consistent $\alpha$ decay chains for $^{288-293}$119
Findings on $\alpha-$decay chain suggest that the nuclides $^{284-296}$119 
have $\alpha-$ decay chain with the life time of the order of micro- or 
nano-second and thus these isotopes might be observed in the laboratory 
through alpha decay. 
 Moreover, the SF half life and alpha decay half life for the isotopic chain $^{284-296}$119 are estimated by the formula of Santosh et al and by CPPM respectively. These investigations showed that the isotopic chain $^{288-293}$119 exhibit 6 $\alpha$ consistent chains followed by SF, the isotopes $^{295,296}$119 undergo 4 $\alpha$ decays followed by SF and the rest of the isotopes undergo consistent alpha decays. Both axially deformed relativistic mean field model and CPPM by Santosh et al
 predict 6 $\alpha$ consistent chains for the  isotopic chain $^{288-293}$119 and thus are in very good agreement with each other.
 
 The isotopes $^{288-293}$119 and $^{295,296}$119 may prove to be of significant interest to experimentalists for future studies.
Thus, we hope that these predictions on the possible decay modes of $Z=119$ superheavy nuclei 
might prove to be quite useful and may serve as a significant input for 
future experimental investigations. 

\section{Acknowledgments}
One of the authors Asloob A. Rather would like to acknowledge Dashty T. Akrawy, Researcher at
Physics Department, College of Science, Salahaddin University, Erbil 44001, Kurdistan, Iraq
 for careful reading of the manuscript and constructive criticism. Asloob A. Rather will also like to express his gratitude to Dorin N. Poenaru, Senior Scientist at Horia Hulubei National Institute of Physics and Nuclear Engineering (IFIN-HH), P.O. Box MG-6, RO-077125 Bucharest-Magurele, Romania 
 for his valuable comments on the write-up and encouragement.


\section*{References}

\end{document}